\documentclass[prd,10pt,twocolumn,showpacs,nofootinbib,aps,superscriptaddress,preprintnumbers]{revtex4-1}

\usepackage{bm}
\usepackage{graphicx}
\usepackage{multirow}
\usepackage{amsmath}
\usepackage{mathrsfs}
\usepackage{amssymb}
\usepackage{hyperref}
\usepackage{yfonts}
\usepackage{color}
\usepackage{xspace}
\usepackage{verbatim}
\usepackage{mathtools}
\usepackage{booktabs}
\usepackage{tabularx}
\usepackage[utf8x]{inputenc} 
\usepackage[dvipsnames,table]{xcolor}
\hypersetup{urlcolor=RedViolet,
	    citecolor=ForestGreen ,
	    linkcolor=RoyalBlue, 
	    colorlinks=true}
\definecolor{lightgray}{rgb}{0.9,0.9,0.9}	    
\definecolor{green}{rgb}{0,0.5,0}
\definecolor{red}{rgb}{1,0,0}
\definecolor{blue}{rgb}{0,0,0.5}

\newcommand{\epsa}{\hat{\boldsymbol{\epsilon}}_1}
\newcommand{\epsb}{\hat{\boldsymbol{\epsilon}}_2}
\newcommand{\vstr}{\mathbf{v}_\mathrm{str}}
\newcommand{\qhat}{\hat{\mathbf{q}}}
\newcommand{\vlab}{\mathbf{v}_\mathrm{lab}}
\newcommand{\vmin}{v_{\rm min}}

\newcommand{\vesc}{v_\mathrm{esc}}

\newcommand{\kms}{\textrm{ km s}^{-1}}

\newcommand{\dbd}[2]{\ifmmode \frac{\textrm{d}#1}{\textrm{d}#2}\else $\textrm{d}#1/\textrm{d}#2$\fi}
\newcommand{\pbp}[2]{\ifmmode \frac{\partial#1}{\partial#2}\else $\partial#1/\partial#2$\fi}
\newcommand{\erf}{\mathrm{erf}}

\DeclareMathAlphabet{\mathpzc}{OT1}{pzc}{m}{it}

\newcommand{\be}{\begin{equation}}
\newcommand{\ee}{\end{equation}}
\newcommand{\bea}{\begin{eqnarray}}
\newcommand{\eea}{\end{eqnarray}}

\begin{document}

\title{A Dark Matter Hurricane: Measuring the S1 Stream with Dark
  Matter Detectors}

\author{Ciaran A. J. O'Hare} \email{ciaran.aj.ohare@gmail.com}
\affiliation{Departamento de F\'isica Te\'orica, Universidad de
  Zaragoza, Pedro Cerbuna 12, E-50009, Zaragoza, Espa\~{n}a}

\author{Christopher McCabe} \email{christopher.mccabe@kcl.ac.uk}
\affiliation{Department of Physics, King's College London, Strand,
  London, WC2R 2LS, United Kingdom}

\author{N. Wyn Evans} \email{nwe@ast.cam.ac.uk}
\affiliation{Institute of Astronomy, Madingley Rd, Cambridge, CB3 0HA, United Kingdom}

\author{GyuChul Myeong}
\affiliation{Institute of Astronomy, Madingley Rd, Cambridge, CB3 0HA, United Kingdom}

\author{Vasily Belokurov}
\affiliation{Institute of Astronomy, Madingley Rd, Cambridge, CB3 0HA, United Kingdom}

\preprint{KCL-PH-TH-2018-38}

\date{\today}
\smallskip
\begin{abstract}
The recently discovered S1 stream passes through the Solar
neighbourhood on a low inclination, counter-rotating orbit. The
progenitor of S1 is a dwarf galaxy with a total mass
comparable to the present-day
Fornax dwarf spheroidal, 
so the stream is expected to have a significant DM component.
We compute the effects of the S1 stream on
WIMP and axion detectors as a function
of the density of its unmeasured dark component. 
In WIMP detectors the S1 stream supplies more high energy nuclear 
recoils so will marginally improve DM detection prospects.
We find that even if S1 comprises less than 10\% of the local density, multi-ton xenon WIMP detectors can distinguish the S1 stream from the bulk halo
in the relatively narrow mass range between~$5$ and~$25$~GeV.
In directional WIMP detectors such as CYGNUS, S1 increases DM detection prospects more substantially 
since it enhances the anisotropy of the WIMP signal.
Finally, we show that axion haloscopes possess by far the greatest potential sensitivity to the S1 stream if its dark matter component is sufficiently cold. 
Once the axion mass has been discovered, the distinctive velocity distribution of S1 can easily be extracted from the axion power spectrum.
\end{abstract}

\maketitle

\section{Introduction}

Dark matter (DM) halos contain a plethora of substructure due to the
tidal disruption and stripping of satellite galaxies or dark subhalos
of the Milky Way (MW). The accretion of material can give rise to
prominent streams of DM particles wrapping the
galaxy. Streams are seen generically in simulations of halos and have
been observed locally in the MW and in nearby
galaxies~\cite{Belokurov:2006,Myeong:2018,Shipp:2018yce}. Such
substructure, being highly kinematically localised, poses excellent
prospects for the direct detection of DM. Hence there is a
sizeable literature on the subject of streams and their signals in
direct detection experiments, see
e.g.\ Refs.~\cite{Lee:2012pf,O'Hare:2014oxa,Savage:2006qr,Foster:2017hbq,OHare:2017yze,Kavanagh:2016xfi}. Historically,
the stream from the Sagittarius dwarf galaxy was used to motivate much
of this
work~\cite{Newberg:2003cu,Yanny:2003zu,Majewski:2003ux,Luque:2016nsz,Purcell:2012sh,Purcell:2011nf}. However,
the last decade has seen the branches of the Sagittarius stream mapped
out in a number of stellar tracers (main sequence turn-off stars, blue
horizontal branch stars, RR Lyrae) across $360^\circ$ on the sky. We
now know that the Sagittarius stream does not pass close to the Sun
~\cite{Koposov:2012, Belokurov:2014} and so it will not have have any
impact on laboratory experiments for the direct detection of dark
matter.  Nonetheless, formalisms developed with the Sagittarius stream
in mind will be useful here.

Many stellar streams have been detected as overdensities of resolved
stars against the background (see~\citet{Newberg:2016} for
reviews). However, there is a much more powerful method of detection
that will enable the identification of $\sim 100-200$ streams in the
inner halo of the MW over the next few years.  Streams remain
kinematically cold and are identifiable as substructure in phase space
long after they have ceased to be recognisable in star counts against
the stellar background of the galaxy. The arrival of the first data
releases from the {\it Gaia} satellite is transformational for our
understanding of substructure in the stellar halo. {\it Gaia} is an
astrometric satellite that is providing distances and proper motions
for over a billion stars in the Galaxy~\citep{Gaia:2016}. When
cross-matched against spectroscopic surveys we can obtain six
dimensional phase space coordinates for MW stars. This enables
searches for comoving groups of stars to be conducted directly in
phase space, and the calculation of statistical measures of
substructure~\citep{Helmi:2017,Myeong:2017,Lancaster:2018}.

\begin{figure}[!t]
 \includegraphics[width=0.49\textwidth,trim={1cm 0cm 0cm 0cm},clip]{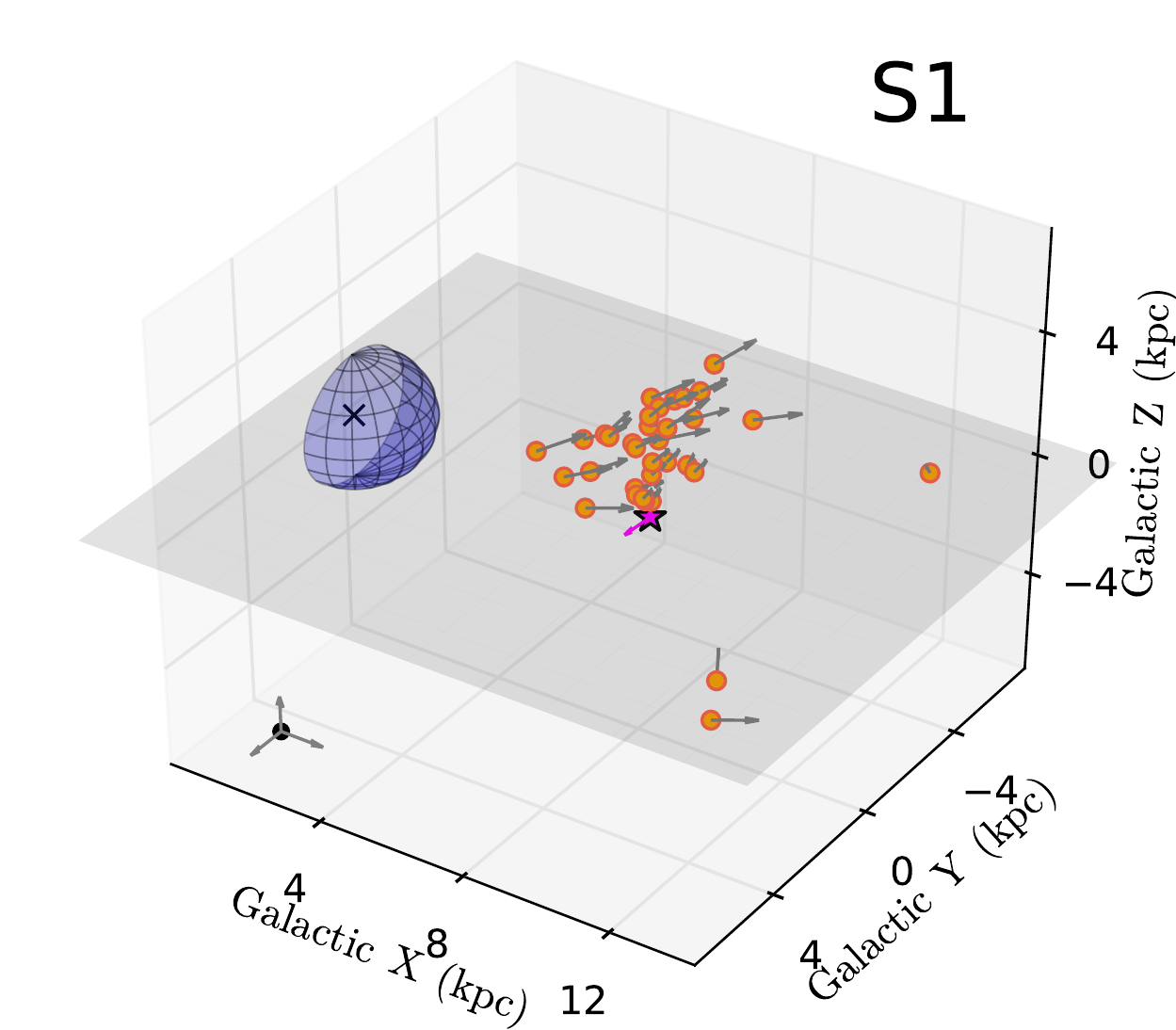}
\caption{The S1 stream in Galactic coordinates with ($X,Y$) defining
  the Galactic plane and $Z$ the height above the disk.  This view is
  partial as it is limited by the footprint of the SDSS-{\it Gaia}
  dataset, whilst the S1 stream extends well beyond the footprint. The
  arrows show the total Galactocentric velocity of the S1 stars. The
  Sun and the Sun's motion are marked as a star and a magenta
  arrow. Notice that the Sun lies in the path of the counter-rotating
  S1 stars.  A $2$\,kpc radius sphere and a grey plane are crude
  representation of the Galactic bulge and the Galactic plane to give
  a sense of S1's size and morphology. A triad of velocity vectors
  (scale of $300$\,km\,s$^{-1}$) is marked in the bottom of each
  panel to illustrate the velocity scale.}
\label{fig:S1pic}
\end{figure}

Here, we draw attention to a remarkable new stream, S1, recently
discovered in data from the Sloan Digital Sky Survey (SDSS) and the
{\it Gaia} satellite~\cite{Myeong:2018,Myeong:Preprint}. The stars in S1 impact on the Solar system at
very high speed almost head-on. A coherent stream of DM
associated with S1 hits the Solar system slap in the face.  The
effects of such a low inclination, retrograde stream are different
from the previously considered, almost polar Sagittarius stream.
Streams can impact the detection of any DM particle to some extent, so
we study the effects of the S1 stream on experiments attempting to
discover candidate particles from light axions or axion like particles (ALPs, $m_a \approx 10^{-10} - 10^{-3}$
eV) to standard weakly interacting massive particles (WIMPs, $m_\chi \gtrsim 1$ GeV). Additionally, since it is well known
that nuclear recoil-based direct WIMP searches possess limited sensitivity
to the DM velocity distribution,\footnote{See the extensive literature accounting 
for astrophysical uncertainty in the analysis of direct detection data~\cite{Peter:2009ak,McCabe:2010zh,Fox:2010bz,Fox:2010bu,Fox:2010bu,Peter:2011eu,McCabe:2011sr,Frandsen:2011gi,Gondolo:2012rs,Kavanagh:2013wba,Kavanagh:2013eya,DelNobile:2013cta,Fox:2014kua,Feldstein:2014gza}. 
Refs.~\cite{Gelmini:2016pei,Gondolo:2017jro,Gelmini:2017aqe,Ibarra:2017mzt,Fowlie:2017ufs,Ibarra:2018yxq} provide the most up to date developments.} we also study the impact of S1 on 
\emph{directional} detectors in which more kinematic information is preserved~\cite{Morgan:2004ys,Billard:2009mf,Lee:2012pf,O'Hare:2014oxa,Mayet:2016zxu,Kavanagh:2016xfi}.

We begin by summarising the observational data on the S1 stream in
Sec.~\ref{sec:S1}. The properties of the halo and stream model are
discussed in Sec.~\ref{sec:halomodel}, while the consequences for 
xenon direct detection experiments, a future directional WIMP detector 
and axion haloscopes are examined in
Secs.~\ref{sec:xenon}-\ref{sec:axions}, respectively. We sum up our
results in Sec.~\ref{sec:summary}.

\begin{figure*}
\includegraphics[width=\textwidth]{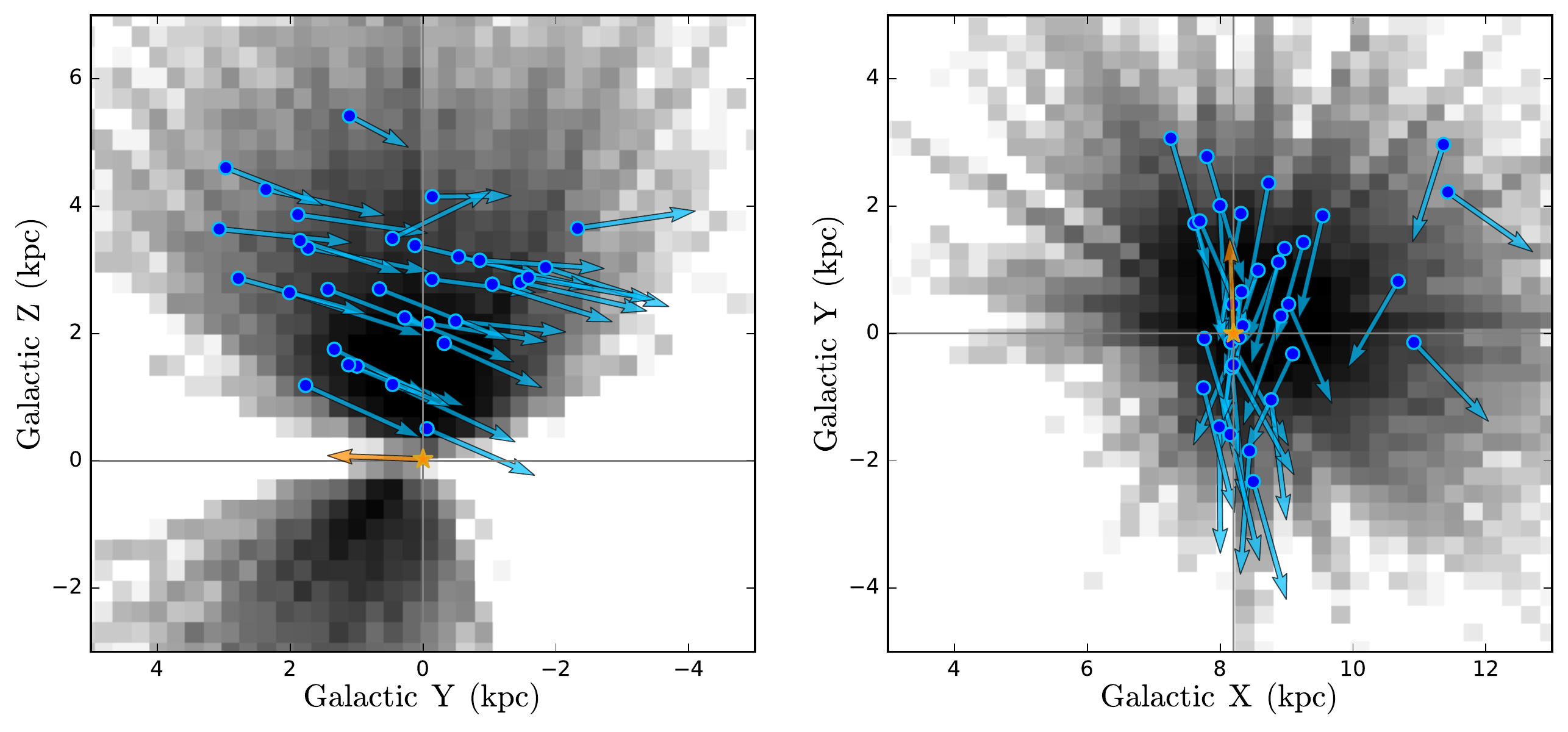}
\caption{The S1 stars projected into the ($Y,Z$) and ($X,Y$) planes.
  The stream is seen sideways-on (left) and face-on (right) in the two
  projections. The Sun's velocity is marked as a yellow arrow, whilst
  the position of the Sun is indicated by the grey crosshair.  S1 has
  modest inclination with respect to the Galactic plane, but it is
  broad ($\sim 2$ kpc) as befits its dwarf galaxy origin. The 34 S1
  stars plotted here were found in searches through the comparatively
  local SDSS-{\it Gaia} dataset~\cite{Myeong:Preprint} shown as a grey distribution, but the full extent
  and morphology of the stream awaits searches through the more
  extensive {\it Gaia} Data Release 2.}
\label{fig:S1picnew}
\end{figure*}

\section{The S1 Stream}\label{sec:S1}

Three-dimensional and projected views of the S1 stream are shown in
Figs.~\ref{fig:S1pic} and~\ref{fig:S1picnew}. It has a low
inclination to the Galactic plane and passes through the Solar
neighbourhood with a velocity that opposes the direction of Galactic
rotation.  S1 was originally discovered in the SDSS-{\it Gaia}
catalogue by searches through 62,133 main-sequence turn-off halo stars
with photometric parallaxes, line of sight velocities, proper motions
and metallicities~\citep{Myeong:2018}. By matching the kinematics of
the S1 stars to numerical libraries of accreted remnants, the
progenitor of the S1 stream is believed to have had a total mass
(stars plus DM) of approximately~$10^{10} M_\odot$ and an infall time of
$\gtrsim 9$~Gyr. The progenitor is comparable to (though somewhat more
massive than) the present-day Fornax galaxy, the largest surviving
dwarf spheroidal in the halo of the MW. If this prediction is true
then the S1 stellar stream must be accompanied by a substantial DM
stream.

The most efficient way to search for substructure is through action
space, rather than velocity space. Actions are adiabatic invariants so
are conserved under slow evolution of the potential. Searches through
the SDSS-{\it Gaia} data in action space revealed a clearer view of
the S1 stream with fewer outliers and
contaminants~\citep{Myeong:Preprint}.  It contains 34 confirmed
stellar members, as shown in Fig.~\ref{fig:S1picnew}. Our view of the
S1 stream is limited by the footprint of the SDSS-{\it Gaia}
survey. The means of the Galactic positions of the stars are $(X,Y,Z)
= (8.9,0.6,2.5)$ kpc, together with dispersions ($1.6, 1.4, 1.9$)~kpc.
The Solar position is (8.2, 0, 0.014) kpc~\cite{Binney:1997,
  McMillan:2017}, so the S1 stellar stream, together with its DM
appendage, is passing directly through the Solar neighbourhood. This
is consistent with analyses of numerical
simulations~\citep{Stiff:2001}, which suggest that there is
$\mathcal{O}(1)$ probability of a substantial DM stream
locally. Though the local halo is smooth in dissipationless
simulations~\citep{Vogelsberger:2008qb}, the existence of stellar
streams, and their accompanying dark matter, shows the importance of
comparatively recent infall and accretion.

The kinematics of the S1 stream make it ideal for DM detection
experiments, as the signature is very different from typical halo
stars.  S1 is counter-rotating with mean velocity $\vstr = (8.6,
-286.7, -67.9)$ km s$^{-1}$. Its velocity dispersion tensor
diagonalised in cylindrical polars is $\sigma_{\rm str} = (115.3,
49.9, 60)$ km s$^{-1}$. The local standard of rest is $v_0 = 232.8$~km\,s$^{-1}$ and the Solar peculiar motion is $(U,V,W) = (11.1, 12.24,
7.25)$ km\,s$^{-1}$ from
Refs.~\cite{Schonrich:2010,McMillan:2017}. Therefore, DM particles
associated with the S1 stream meet the Solar system with a huge
relative velocity, primarily directed along the stream. By contrast,
DM particles in the halo are expected to have a roughly isotropic and
Maxwellian velocity distribution. The S1 stream is more akin to a 
`hurricane' compared to the DM `wind' associated with the halo.

The S1 stellar stream is broad with a width of $\sim 2$~kpc. There
also appears to be a surviving globular cluster (NGC 3201) associated with the
stream which resided in the progenitor galaxy. These facts
corroborate the original suggestion of Ref.~\citep{Myeong:2018} that
the progenitor was a massive dwarf spheroidal, substantial enough to
contain its own retinue of globular clusters. Taking our cue from the
largest dwarf spheroidals like Fornax, a mass-to-light ratio of $\sim
10-100$ seems very realistic (see Table 5 of
Ref.~\cite{Amorisco:2011}). The density of the DM component cannot
easily be measured from the stellar stream. However, follow up studies with the \emph{Gaia} Data
Release 2~\cite{GaiaDR2} will see S1 traced throughout the Galaxy, in particular for lower values of $Z$ than shown 
in Fig.~\ref{fig:S1picnew}. This will provide improved constraints on
its contribution to the local DM density from modelling of
the disruption of the progenitor. For the moment, we wish to know how
dense the stream must be if it is to be detected in an ongoing or
future terrestrial direct detection experiment.

\section{Models of the Halo and S1 Stream}\label{sec:halomodel}

We can model both the smooth component of the DM halo and the S1 stream
with a Maxwellian boosted by some velocity. For the the smooth
component, we assume the local speed distribution from the standard
halo model (SHM) and boost by the lab velocity,
\begin{align}\label{eq:shm}
f_\mathrm{SHM}(\mathbf{v},t) = \frac{1}{(2\pi
  \sigma_v^2)^{3/2}N_\mathrm{esc}} \, \exp \left( - \frac{|\mathbf{v}
  - \mathbf{v}_\textrm{lab}(t)|^2}{2\sigma_v^2}\right) \,
\nonumber\\ \times \Theta (\vesc - |\mathbf{v} -
\mathbf{v}_\textrm{lab}(t)|)\,,
\end{align}
where the constant $N_{\rm esc}$ is used to renormalise the distribution 
after truncating at the local escape velocity $\vesc$ using the
Heaviside function $\Theta$, namely:
\begin{equation}
N_\mathrm{esc} = \erf \left( \frac{\vesc}{\sqrt{2}\sigma_v}\right) -
\sqrt{\frac{2}{\pi}} \frac{\vesc}{\sigma_v} \exp \left(
-\frac{\vesc^2}{2\sigma_v^2} \right)\,.
\end{equation}
The escape speed locally is $\approx 520$ km s$^{-1}$
~\cite{Williams:2017} and in the SHM, $\sqrt{2} \sigma_v = v_0$, where
$v_0$ is the amplitude of the rotation curve and $\sigma_v$ is the one-dimensional dispersion velocity. To make a fair
comparison of null results in different experiments a benchmark halo
model is needed. The SHM is widely used on account of its simplicity
and flexibility, but many elaborations are possible. For example, the
effects of triaxiality, velocity anisotropy and dark substructures
have all received attention in the context of direct DM detection
before~\cite{Ca00,Savage:2006qr,Bruch:2008rx,MarchRussell:2008dy,Lisanti:2010qx,Lisanti:2011as,Lee:2012pf,Vergados:2012xn,Billard:2012qu,O'Hare:2014oxa,Foster:2017hbq,OHare:2017yze,Fornasa:2013iaa,Bozorgnia:2013pua,Kavanagh:2016xfi}. 
Furthermore, recent analyses using hydrodynamic simulations and astrometric data
have suggested that the DM halo may be colder than assumed
here~\cite{Bozorgnia:2016ogo,Lentz:2017aay,Sloane:2016kyi,Herzog-Arbeitman:2017zbm,Necib:2018iwb}. Here
we merely remark that if this is the case then this would emphasise
the presence of S1.

Assuming the velocity distribution of a stream is also Maxwellian, we
only need to make the replacements $\sigma_v \rightarrow \sigma_{\rm
  str}$ and $\vlab(t) \rightarrow \vlab(t) - \vstr$.  Since the
stellar stream seems to be somewhat anisotropic, we can model its
velocity distribution by generalising the isotropic Maxwellian
introduced above,
\begin{align}
& f_\mathrm{str}(\mathbf{v}, t) = \frac{1}{(8 \pi^3 \det{\boldsymbol{\sigma}^2})^{1/2}}  \times \\&\exp\left(-(\mathbf{v} - \vlab(t)+\vstr)^T \frac{\boldsymbol{\sigma}^{-2}}{2} (\mathbf{v} - \vlab(t)+\vstr) \right) \nonumber \, .
\end{align}
To write this formula in an analytic form, we have here ignored the
truncation at the local escape speed. In our numerical work, we
include the truncation even though the correction is small. For the S1
stream we can assume that the dispersion tensor is diagonal
$\boldsymbol{\sigma}^2 = \textrm{diag}(\sigma^2_r, \sigma^2_\phi,
\sigma^2_z)$. We use the stellar dispersion tensor derived in cylindrical coordinates,
 however for the local distribution sampled on Earth the distinction 
 between cylindrical and spherical polars is negligible. 

To combine the stream with an isotropic halo model we assume that it
comprises some fraction of the local density $\rho_{\rm str}/\rho_0$,
so that the total distribution is
\begin{equation}\label{eq:SHMStr}
 f_\mathrm{SHM+str}(\mathbf{v}) =
 \left(1-\frac{\rho_\mathrm{str}}{\rho_0}\right)f_\mathrm{SHM}(\mathbf{v},t)
 + \frac{\rho_\mathrm{str}}{\rho_0} f_\mathrm{str}(\mathbf{v},t) \, .
\end{equation}
Although $\rho_0=0.3$ GeV cm$^{-3}$ is a widely-used value of the
local DM density, more recent investigations using
vertical kinematics of stars tend to find the somewhat larger value of
$\rho_0 \approx 0.5$ GeV cm$^{-3}$ \cite{Smith:2012,Bienayme:2014,Sivertsson:2017rkp,Buch:2018qdr}.

Of course, the underlying assumption here is that the DM particles
have the same kinematic properties as the stars. This is unlikely to
be correct in detail.  For example, the DM streams of Sagittarius are
believed to be more extended then the stellar streams and misaligned
from them~\citep{Purcell:2012sh,Purcell:2011nf}. Judging from the mass
of its stellar content, the Sagittarius progenitor is almost certainly
a dwarf irregular galaxy~\citep{NO10}, whereas the S1 progenitor is a
dwarf spheroidal~\citep{Myeong:2018}. In the former case, the stars
are distributed in a disk, whereas the DM is spheroidal, so mismatches
between the stellar and DM tails are only to be expected. In the
latter case, the stars and DM start out as both spheroidally
distributed, though possibly with different flattenings. The process
of tidal stripping does refashion the more compact stellar and more
extended DM content differently~\citep[e.g.,][]{Sa18}, so mismatches
are still possible -- but perhaps not as substantial as in the case of
dwarf irregulars. Similarly, the velocity dispersion of DM particles
in dwarf spheroidals is somewhat larger then the dispersion of the
stars~\citep{Am11} -- against which must be balanced the fact there
almost certainly remain some contaminants in the S1 stars, so our
present stellar dispersion may be an overestimate. In fact, the
velocity dispersion of a stream can evolve considerably both with time
since disruption and along the stream at the present
day~\citep{Er16,Gi17}.

\begin{figure}[t]
\includegraphics[width=0.49\textwidth]{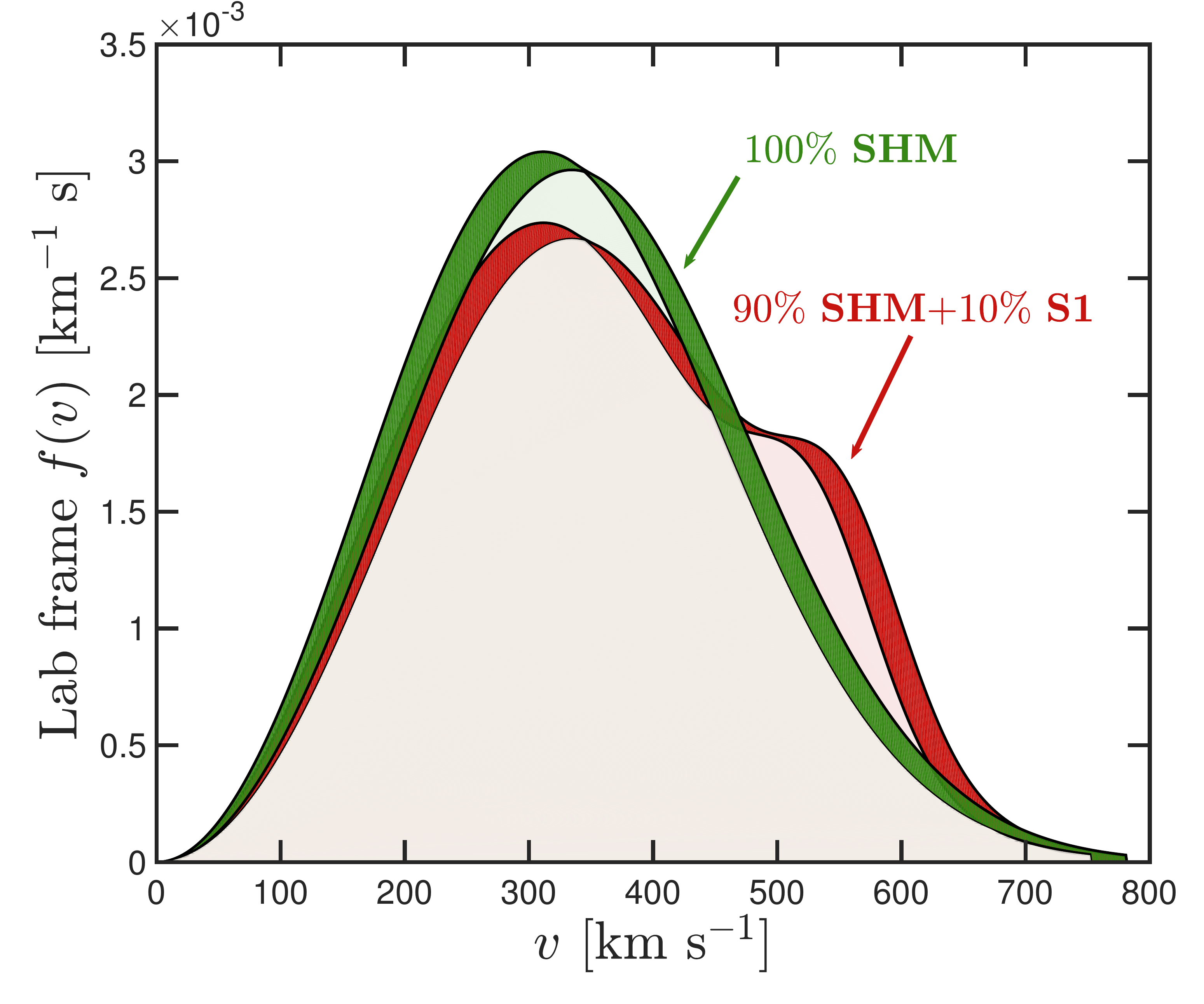}
\caption{Laboratory frame speed distributions for the SHM (green) and SHM+S1 (red) models. The shaded region delimits the range taken by the speed distribution modulated over one year. In the SHM+S1 model we have assumed that the stream comprises 10\% of $\rho_0$.}\label{fig:fv}
\end{figure}

We show the range of $f(v)$ in the lab frame (which is modulated over
one year) in Fig.~\ref{fig:fv} for both the SHM and SHM+S1 model,
assuming $\rho_{\rm str}/\rho_0 = 0.1$.  We clearly see that the
SHM+S1 model has a larger number of high speed DM particles compared
to the SHM alone.  The distribution in this case was calculated by
numerically integrating the 3-dimensional multivariate Gaussian form
for $f(\mathbf{v})$ including dispersion velocities $\sigma_{\rm
  str}^{r,\phi,z}$ in each direction. All the results we present here
are essentially insensitive to this multivariate treatment of the
stream velocity distribution. One could instead use, more
straightforwardly, the same velocity dispersion in all three
directions (for which there are analytic formulae for all necessary
direct detection signals). Accounting for the annual modulation, the
average value that best reproduces the full multivariate distribution
is $\sigma_{\rm str} \approx 46$~km~s$^{-1}$.

The velocity of the lab (and hence the lab frame velocity of the
stream) is time dependent due to the revolution and rotation of the
Earth. This gives rise to well known annual and diurnal
modulations~\cite{Drukier:1986tm, Freese:1987wu}. The diurnal
modulation in speed is likely unobservable for any realistic
experiment (with the possible exception of certain axion
experiments~\cite{Knirck:2018knd}), so we focus on the annual
effect. We calculate the velocity of the lab using formulae detailed
in Ref.~\cite{McCabe:2013kea, Mayet:2016zxu}. The velocity of the Sun
is set by the velocity of the local standard of rest and the peculiar
velocity of the Sun with respect to the LSR: $\textbf{v}_\odot =
(11.1,232.8+12.24,7.25)$~km~s$^{-1}$. When combined with the Earth
revolution velocity, for the year 2018 we find
\begin{equation}
\vlab = \mathbf{v}_\odot + v_\oplus \left( \cos[\omega(t-t_a)]\, \epsa + \sin[\omega (t-t_a)] \,\epsb \right)
\end{equation}
where $\omega = 2\pi/(365\,{\rm days})$, $t_a = $ 22 March, 
$v_\oplus = 29.79~\mathrm{km}~\mathrm{s}^{-1}$ and the vectors are,
\begin{align}
\epsa &= (0.9941, 0.1088, 0.0042)^T  \, , \\ 
\epsb &= (-0.0504, 0.4946, -0.8677)^T \, .
\end{align}

We emphasise again that our assumptions made for the various input
astrophysical parameters are a departure from the commonly agreed upon
benchmarks. Here we favour instead more recent determinations, notably
$\rho_0 = 0.5$ GeV cm$^{-3}$, $\vesc = 520$ km s$^{-1}$ and $v_0 =
232.8$ km s$^{-1}$. This is in part to obtain some self-consistency
given that we are using a particular determination of the stream
velocity. In addition it enables us to advertise the ongoing
refinement of these values.

\section{Sensitivity of Xenon detectors}\label{sec:xenon}

Current and existing dual phase xenon detectors~\cite{Chepel:2012sj} are the most
sensitive to DM-induced nuclear recoils for WIMP DM that has a mass $m_\chi \gtrsim 5$~GeV.
The rate~$R$ of spin independent (SI) nuclear scattering is expressed as a function of the nucleus'
recoil energy~$E_r$,
\begin{equation}\label{eq:finaleventrate}
 \frac{\textrm{d}R(t)}{\textrm{d}E_r} = \frac{\rho_0}{2\mu_{\chi p}^2 m_\chi} \,\sigma^{\rm SI}_p \, \mathcal{C}_\textrm{SI} \,F^2(E_r)  \, g(\vmin,t) \, ,
\end{equation}
where $\mu_{\chi p}$ is the WIMP-proton reduced mass and $\sigma^{\rm SI}_p$ is the WIMP-proton scattering cross-section. In this formula,
we have absorbed all the dependence on the nuclear content into a
form factor~$F(E_r)$, for which we use the Helm parameterisation~\cite{Vietze:2014vsa}, and an `enhancement factor'~$\mathcal{C}_{\rm SI}$. 
For a nucleus with mass number $A$ and atomic number $Z$, and assigning
the couplings to neutrons and protons $f_n$ and $f_p$, the enhancement factor~is
 \begin{equation}\label{eq:CSI}
 \mathcal{C}_{\rm SI} = | Z + (f_n/f_p)(A-Z)|^2 \, .
\end{equation}
We assume equal couplings to protons and neutrons, $f_n/f_p = 1$, as
generically found in models with a Higgs-like mediator~\cite{Hoferichter:2017olk},
though different values are possible in other models (see e.g.~\cite{Frandsen:2011cg}).

The most important function for the purposes of this study is $g(\vmin,t)$,
which contains all of the dependence on the DM velocity distribution:
 \begin{equation}\label{eq:gvmin}
g(v_{\rm min},t) = \int_{v>\vmin}^\infty \frac{f(\textbf{v},t)}{v} \, \textrm{d}^3 v \, .
 \end{equation}
 Here, $f(\textbf{v},t)$ is the DM velocity distribution in the lab frame. For the smooth, isotropic DM halo model,
 we use the distribution in Eq.~\eqref{eq:shm}, while for the SHM+S1 model, we use the distribution in Eq.~\eqref{eq:SHMStr}
 and treat $\rho_{\rm str}/\rho_0$ as a free parameter.
 
Physically, $g(\vmin,t)$ is the mean inverse speed for particles 
that have a speed greater than $\vmin$, the minimum DM speed 
for which the nucleus recoils with energy $E_r$. Simple kinematics 
results in the relation $\vmin = \sqrt{m_N E_r/ (2 \mu^2_{\chi N})}$, where $m_N$ is the xenon nucleus mass.

\begin{figure}[t]
\includegraphics[width=0.49\textwidth]{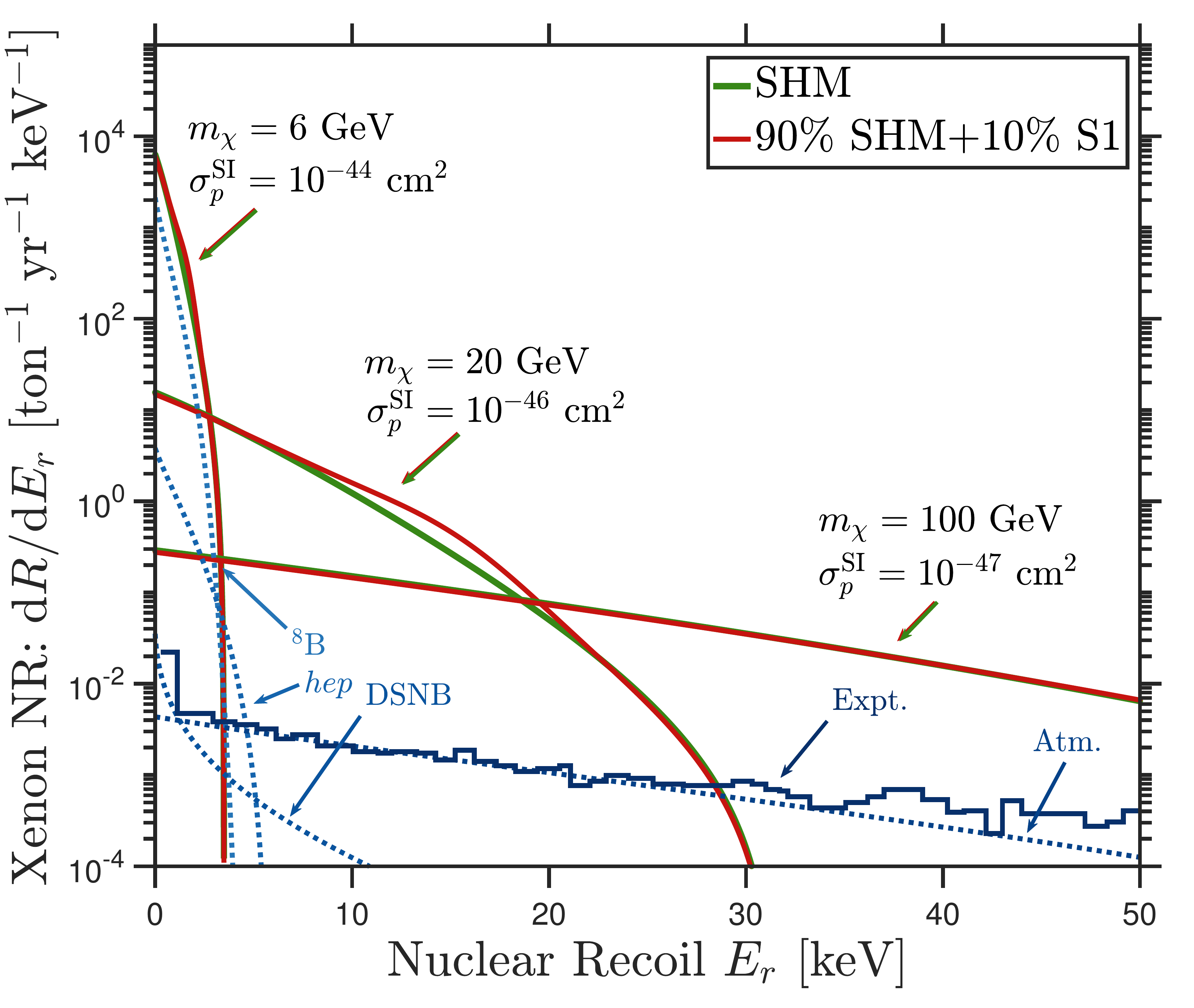}
\caption{Differential xenon recoil spectra as a function of nuclear recoil energy for DM with a mass 6, 20 and 100 GeV. For each mass we show both the recoil distribution for the SHM (green) and the SHM with a 10\% contribution from S1 (red). The cross section in each case is chosen for illustrative purposes and lie near to current exclusion limits. Except for the 20 GeV spectra, the red and green lines are nearly indistinguishable. In blue we show the main nuclear recoil backgrounds. These include the four neutrino backgrounds ($^8$B, $hep$, diffuse supernova and atmospheric) as well as the detector and environmental background in LZ, labelled `Expt.', which we take as a proxy for all xenon detectors. The spectra displayed here do not include the effects of energy resolution and detection efficiency.}\label{fig:xenon-rate}
\end{figure}

 We show the differential event rates for three WIMP masses under both the SHM and SHM+S1 models in Fig.~\ref{fig:xenon-rate}. 
 Most notably we see that for the largest mass displayed here (100~GeV), the two models look essentially identical
while at the smallest mass shown (6~GeV), there is a small difference. It is only for the intermediate mass (20~GeV)
where the two spectra are easily distinguishable. The main feature provided by S1 is an excess for recoil energies where $v_{\rm min}(E_r) \lesssim |\mathbf{v}_{\rm lab} - \mathbf{v}_{\rm str}| \approx 550 \kms$ above which the event rate then decreases back down to SHM-only case. This is because $g(v_{\rm min})$ for a stream is essentially a step function but with a smooth rather than a sharp cutoff due to the stream dispersion. 

We can understand why the two distributions become difficult to distinguish at large masses by considering the range of speeds an experiment is sensitive to for a given WIMP mass. The lower limit of this range is bounded by the threshold of the experiment $v_{\rm min}(E_{\rm th})$ and the upper limit is given approximately by the point at which the form factor is suppressing most of the event rate $v_{\rm min}(E_{\rm max})$. As we increase $m_{\chi}$ the value of $v_{\rm min}$ for a given recoil energy decreases. This means larger masses are sampling lower speeds. So eventually for very large masses the window of speeds which give measurable recoil energies is below the characteristic step of the $g(\vmin)$ for the stream. For these large masses the only thing distinguishing the SHM and SHM+S1 distributions is that the rate appears to be absent of recoils by a fraction $\sim \rho_{\rm str}/\rho_0$ which are missing at higher energies. This subtlety regarding the effects of the stream as a function the WIMP mass will become important when we calculate its discoverability in Sec.~\ref{sec:streamDLs}.
 
 \subsection{Experimental details}\label{subsec:expdetails}
 
In practice the measurement of $\textrm{d}R/\textrm{d}E_r$ will be
hindered by various detector effects and backgrounds. 
The efficiency of nuclear
recoil detection decreases sharply towards recoil energies $E_r\lesssim1$ keV
meaning that the measured spectrum of both the signal and
background will be suppressed below these energies. In this work, we use the LZ
nuclear recoil efficiency curve from Ref.~\cite{Akerib:2018lyp} and assume that it serves as a proxy for all future multi-ton
xenon detectors. For an effective energy resolution we apply a Gaussian smearing over the recoil spectrum with a width $\sigma_E(E_r) =  0.5\, {\rm keV}\,\sqrt{E_r/1\,{\rm keV}}$~\cite{Schumann:2015cpa}. 

An experiment will also see some nuclear recoil events from radioactive 
material in the detector and from the environment, 
as well as events from cosmic and terrestrial neutrinos. All of these sources constitute a background
to the DM signal events. We include the `materials' background calculated for
LZ~\cite{Akerib:2018lyp}, again assuming that it serves as a proxy for all future xenon detectors. 
This approximately has the shape of two exponentially decaying spectra: 
one sharply decaying at low energies and another slowly decaying over higher energies. 
We assume that the shape of the spectrum is known but
parameterise the overall normalisation with a nuisance parameter
$R_{\rm bg}$ with an uncertainty of 20\%~\cite{Akerib:2018lyp}.

For experiments with a multi ton-year exposure, the background from the coherent scattering between
neutrinos and nuclei becomes important. The
details of the background from Solar, diffuse supernova and
atmospheric neutrinos can be found in, for example,
Refs.~\cite{Billard:2013qya,Ruppin:2014bra,OHare:2016pjy}. The most
important neutrino background for xenon experiments are those from the
Solar $^8$B decay. To parameterise the nuclear recoil rate due to this
background and its uncertainty, we look to the determination from the
Solar global analysis of Bergstr\"om et al. $\Phi_{^8{\rm B}} =
5.16^{+0.13}_{-0.09} \times 10^6$ cm$^{-2}$
s$^{-1}$~\cite{Bergstrom:2016cbh}, which currently has a smaller
uncertainty than high or low metallicity Solar model predictions,
e.g.\ Ref.~\cite{Vinyoles:2016djt}.

In a xenon experiment, the shape of the nuclear recoil spectrum from $^8$B neutrinos
 looks remarkably
similar to the spectrum from a 6~GeV WIMP. The similarity, coupled with the fact that the
flux of neutrinos possesses a systematic uncertainty, means that the
background will inhibit the discovery of WIMPs of certain masses like
6~GeV. The limit at which neutrinos begin to cause sub-Poissonian
background rejection occurs for exposures beyond the ton-scale and
imprints on limit projections a shape known as the neutrino floor. In 
Fig.~\ref{fig:xenon-rate} we also show the principal neutrino backgrounds 
that contribute to the neutrino floor in a xenon experiment.

In addition to Solar neutrinos, we also must consider the diffuse
supernova neutrino background (DSNB) from the cosmological history 
of supernova explosions. This has not been measured but calculations predict it to have a low flux,
$\Phi_{\rm DSNB} = 85.7 \pm 42.7$ cm$^{-2}$
s$^{-1}$~\cite{Beacom:2010kk}. Then affecting the discovery of even higher masses
we have the background due to atmospheric neutrinos; like the DSNB its
flux also possesses a large theoretical uncertainty at the relevant
energy tail, $\Phi_{\rm Atm} = 10.54 \pm 2.1$ cm$^{-2}$
s$^{-1}$~\cite{Honda:2011nf}.

\subsection{Statistical test for the presence of S1}

We wish to know how dense the S1 stream must be if it is to be detected in 
an upcoming multi-ton xenon detector. Our tool for quantifying the stream density
required for a detection of S1 is a hypothesis test using the profile likelihood ratio statistic.
This statistical methodology is in
common use for computing exclusion and sensitivity limits for dark
matter experiments (see e.g.~\cite{Billard:2013gfa,Edwards:2017mnf}).

The profile likelihood ratio test compares the SHM+S1 model
$\mathcal{M}_{\mathrm{SHM}+\mathrm{S1}}$
with parameters $(\rho_{\rm str},\Theta)$ against the SHM model, $\mathcal{M}_{\mathrm{SHM}}$ with parameters $(\Theta)$,
where the DM halo is smooth, isotropic and stream-less. 
The two models differ by one parameter $\rho_{\rm str}$. To test for a non-zero value of this parameter in the data we construct the profile likelihood ratio,
\begin{equation}\label{eq:likelihood-ratio}
\Lambda = \frac{ \mathscr{L} (0,\hat{\hat{\Theta}} ) }{\mathscr{L} (\hat{\rho}_{\rm str},\hat{\Theta} )  }\, ,
\end{equation}
where~$\mathscr{L}$ is a likelihood function which is maximised at
$\hat{\hat{\Theta}}$ when $\rho_{\rm str}$ is set to 0, and $(\hat{\rho}_{\rm str},\hat{\Theta})$ when $\rho_{\rm str}$ is a free parameter. We can use this ratio because our stream-less model
$\mathcal{M}_{\mathrm{SHM}}$ is a subset of the more general model
$\mathcal{M}_{\mathrm{SHM}+\mathrm{S1}}$, found after applying the constraint $\rho_{\rm str} = 0$ (cf.~Eq.~\eqref{eq:SHMStr}).

We next define the profile likelihood ratio test statistic,
		\begin{equation}
			q_0 = \left\{ \begin{array}{rl}
			-2\ln \Lambda  & \, \, 0\le \hat{\rho}_{\rm str}\le1 \,,\\
			0  & \, \, \hat{\rho}_{\rm str}<0, \, \, \hat{\rho}_{\rm str}>1 \,.
			\end{array} \right. 
		\end{equation}
According to Wilks' theorem, the
test statistic for the discovery of a signal 
is distributed according to $\frac{1}{2}\chi^2_{1}$ since the two models differ by a constraint applied to one parameter. 
This means that the significance of the signal is obtained from the
simple formula 
$\sqrt{q_0}$.\footnote{For a derivation of this result and a more extensive discussion,
  we refer the reader to Ref.~\cite{Cowan:2010js}.}

In our approach we adopt a binned likelihood for $\mathscr{L}$ so that we can employ the Asimov dataset formalism~\cite{Cowan:2010js}. 
In this formalism, the number of observed events in each bin is set equal to the number of expected events.
The value of the profile likelihood ratio test statistic then asymptotes to the median value that would be obtained from many realisations.
This method saves on expensive Monte Carlo simulations while still allowing accurate results to be obtained.
We have explicitly checked the Asimov dataset formalism against Monte Carlo simulations in a limited range of
parameter space and the agreement is excellent. This is because the detection of the S1 stream requires a large number
of events, so the asymptotic results from Ref.~\cite{Cowan:2010js} are accurate.

\begin{figure*}[t]
\includegraphics[width=0.49\textwidth]{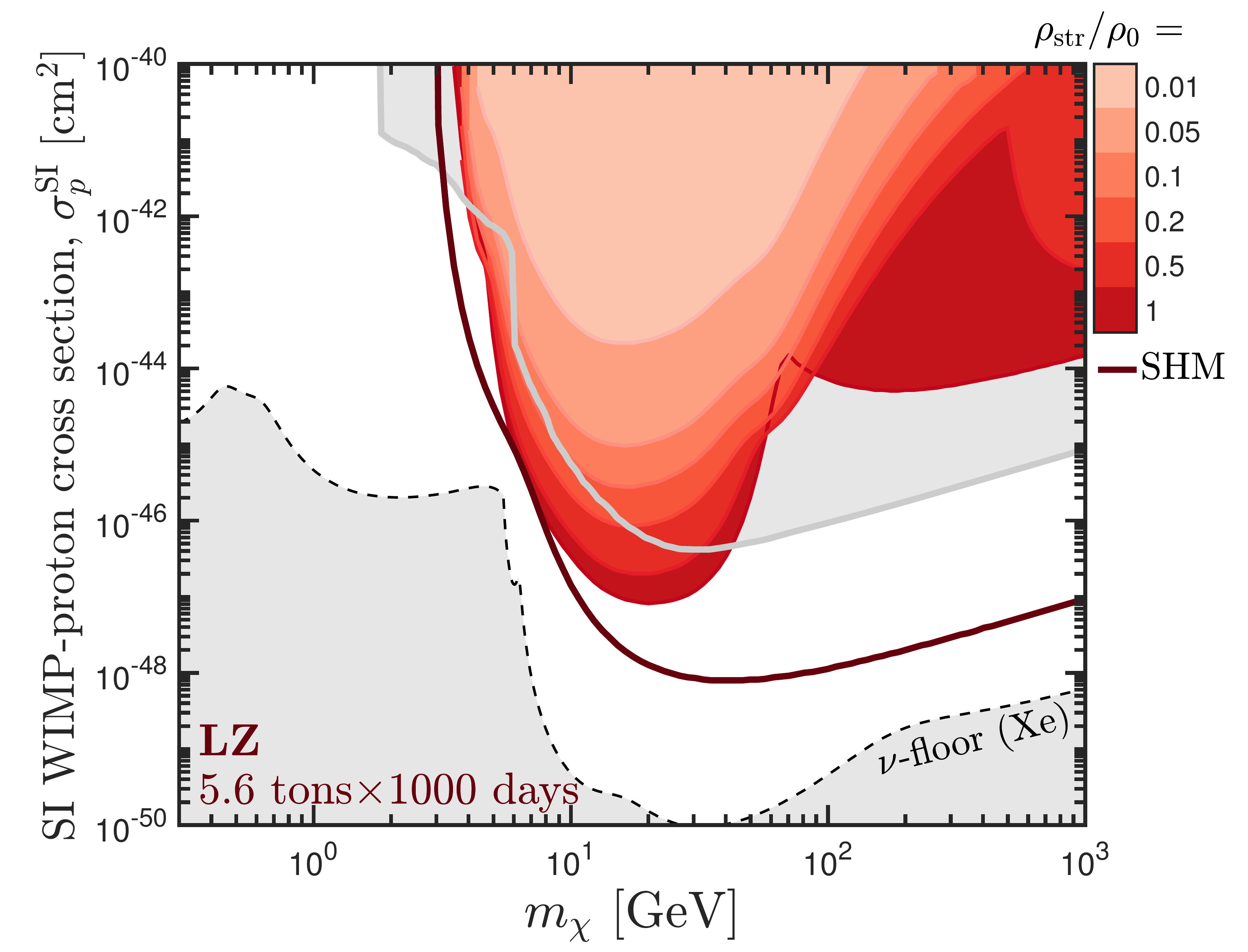}
\includegraphics[width=0.49\textwidth]{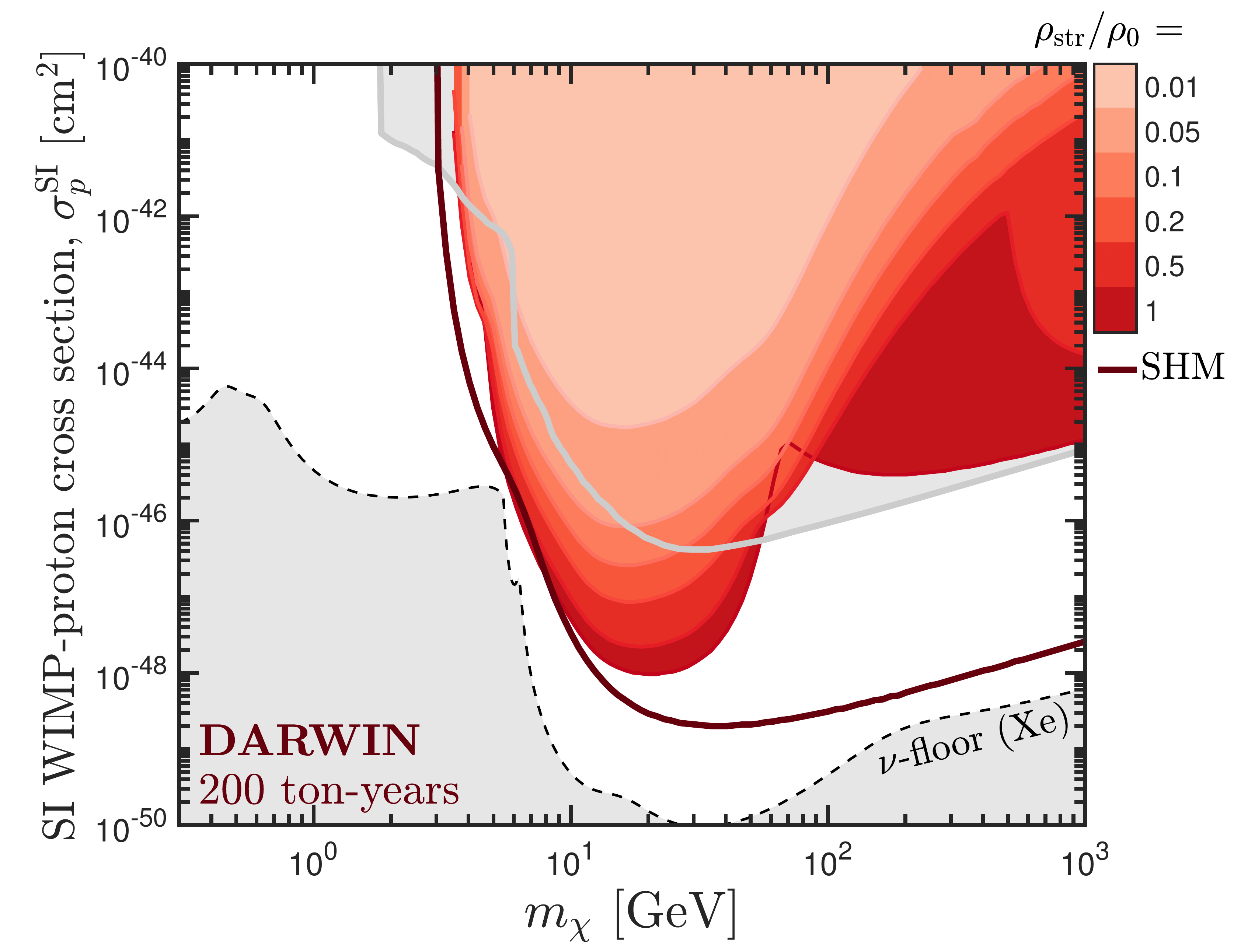}
\caption{Stream discovery limits for {\bf left:} LZ $(5.6~\mathrm{ton}\times1000~\mathrm{day}$ exposure) 
and {\bf right:} DARWIN (200 ton-year
  exposure). The
  coloured shaded regions indicate the values of WIMP masses and cross
  sections required for the median of each experiment to detect S1 at 3$\sigma$ for a given density. The colours from light to dark indicate density fractions from 0.01$\rho_0$ to $\rho_0$. The upper grey boundary in each panel
  shows the cross sections already excluded by experiment. The lower
  dashed line shows the neutrino floor for xenon. If S1 comprises less than 10\% of the local density these detectors can distinguish the S1 stream from the bulk halo in the narrow mass range between~$5$ and~$25$~GeV.
  }\label{fig:streamDL_xenon}
\end{figure*}

The binned likelihood that enters Eq.~\eqref{eq:likelihood-ratio} is the product of the Poisson probability distribution function $\mathscr{P}$ for $N_{\rm obs}$ events, given an expected number of signal and background events $N_{\chi} + \sum N_{\rm bg}$, where the sum extends over all background components. The WIMP and background parameters are shared by both models $\Theta = \{m_\chi$,  $\sigma_p, \mathbf{R}_{\rm bg}\}$. The background is divided further into $\mathbf{R}_{\rm bg} = \{ R_{\rm bg}^1, ..., R_{\rm bg}^{n_{\rm bg}} \}$ used to normalise the $n_{\rm bg}$ background signal rates. In full, the likelihood for $N_{\rm bins}$ in the complete SHM+S1+Background model is
\begin{align}
\label{eq:Like}
\mathscr{L}(&\rho_{\rm str},m_\chi,\sigma_p,\,\mathbf{R}_{\rm bg} ) =\\ &\prod_{i=1}^{ N_{\rm bins}}\mathscr{P}\left[N_\textrm{obs}^i \bigg| N^i_\chi(\rho_{\rm str},m_\chi,\sigma_p) + \sum_{j=1}^{n_{\rm bg}}N^{ij}_{\rm bg}(R^j_{\rm bg}) \right]   \nonumber\\ 
 \times &\prod_{k=1}^{n_{\rm bg}}\mathscr{L}_k (R_{\rm bg}^k) \, .
\end{align}
The likelihood functions $\mathscr{L}_k(R_{\rm bg}^k)$ incorporate the uncertainty for each background component. We assume that the $\mathscr{L}_k$ functions are all Gaussian with the uncertainties discussed in Sec.~\ref{subsec:expdetails}. In all our results we assume $n_{\rm bg} = 5$ backgrounds (4 neutrino and 1 laboratory). 

We utilise a different binning depending on the experiment.  For xenon detectors, $N_{\rm bins}$ is the number of bins in energy. When timing information is included (as in Sec.~\ref{subsec:anmod}), the number of energy bins is multiplied by the number of bins in time. In the case of directional experiments (described in Sec.~\ref{sec:directional}) we multiply by the number of bins in angle as well as in time.

\subsection{S1 discovery limits and their interpretation}\label{sec:streamDLs}

The red shaded regions in Fig.~\ref{fig:streamDL_xenon} show a set of 
S1 stream `discovery limits' for two future xenon experiments
for a range of values of $\rho_{\rm{str}}/\rho_0$.
For a given value of the DM mass, the discovery limit shows the minimum cross section required to
infer the presence of the S1 stream. In this study, for the inference of the S1 stream, we require that the `median' experiment
 can discriminate between the SHM+S1 and SHM models with a significance of 3$\sigma$
(equivalent to $q_0=9$).

The left panel in Fig.~\ref{fig:streamDL_xenon} shows results for LZ~\cite{Mount:2017qzi}, where we assume a
5.6 ton fiducial mass running for 1000 days. A similar sensitivity is
expected from the XENONnT~\cite{Aprile:2017aty} and PandaX~\cite{Cao:2014jsa} detectors, which should
have results in a similar timeframe as LZ.
The right panel in Fig.~\ref{fig:streamDL_xenon} shows results for DARWIN, a hypothetical xenon experiment
that aims to have a significantly larger fiducial mass of around 30~ton~\cite{Aalbers:2016jon}. 
In our results, we assume a total exposure for DARWIN of 200 ton-years.
The upper solid grey lines and upper grey shaded regions in Fig.~\ref{fig:streamDL_xenon} show
the existing exclusion limits on the SI WIMP-proton cross section.
This is an interpolation of the limits of (from low to high masses) CRESST~\cite{Angloher:2015ewa}, 
DarkSide-50~\cite{Agnes:2018oej}, LUX~\cite{Akerib:2016vxi}, PandaX~\cite{Tan:2016zwf} and XENON1T~\cite{Aprile:2018dbl}.
The lower shaded region is the neutrino floor for a xenon target. 
We have recalculated this limit with our choice of astrophysical parameters discussed in Sec.~\ref{sec:halomodel}.
Our calculation follows the procedure introduced in Ref.~\cite{Billard:2013qya} and subsequently used in Refs.~\cite{Ruppin:2014bra,O'Hare:2015mda,Dent:2016iht,Dent:2016wor,Bertuzzo:2017tuf,OHare:2017rag,Gonzalez-Garcia:2018dep,Gelmini:2018ogy,Boehm:2018sux}.

The solid maroon lines show the WIMP discovery limits for the SHM model. 
This line provides a good estimate to the cross section at which a typical experiment 
would first find evidence for DM at a significance of $3\sigma$. Calculating this limit involves the same procedure detailed above, but focused around testing for $\sigma_p$ against a background-only hypothesis, rather than for $\rho_{\rm str}$ against an SHM-only hypothesis. 

As one would expect, to detect the signal \emph{and} to infer the presence of S1 in that signal requires more events than for just seeing WIMP events over a background. This explains why the S1 stream discovery limits all lie above the SHM sensitivity line.
As the S1 density fraction $\rho_{\rm str}/\rho_0$ decreases, more signal events are required
to infer the presence of S1, which means that the DM must interact with a larger scattering
cross-section. This is the behaviour that is demonstrated in Fig.~\ref{fig:streamDL_xenon}.

The prospects for measuring S1 with LZ or DARWIN seem, unfortunately, to be rather limited.
Even for exceedingly large values of $\rho_{\rm str}/\rho_0$, the S1 stream is
unmeasurable above $\sim50$~GeV since the cross section
 to distinguish it from the SHM
lies in the parameter space that has already been excluded.
For more realistic values $\rho_{\rm str}/\rho_0\lesssim10\%$, LZ could 
detect S1 in the narrow DM mass range between approximately 4~and~6 GeV
while DARWIN extends this up to approximately~25~GeV. Towards smaller DM masses,  DARWIN may still be able to detect the S1 stream 
even for small values of $\rho_{\rm str}/\rho_0$. These values of the scattering cross section lie close to existing exclusion 
limits so there would need to be a DM discovery soon for DARWIN to make this detection.

The S1 discovery limits begin to increase sharply above $\sim50$~GeV. Above this mass, as was shown in Fig.~\ref{fig:xenon-rate}, it becomes much harder to discriminate between the SHM+S1 mixed model 
and the pure SHM model because the energies that the experiment is sensitive to are sampling lower speeds where the stream is less prominent. However this trend seems to abruptly stop at a critical mass. This can be seen in Fig.~\ref{fig:streamDL_xenon} as a peak in the 100\% and 50\% stream discovery limits, but is in fact a trend that continues for all smaller values of $\rho_{\rm str}$, but for increasingly large masses, well beyond 1000 GeV. An intuition for this behaviour links back to a subtlety that we highlighted earlier about the shapes of the distributions that we are comparing. For a given stream fraction, there is a characteristic speed at which the SHM only and SHM+S1 distributions cross over (exhibited both in $f(\mathbf{v},t)$ and $g(\vmin,t)$). This means that there is a critical value of $m_\chi$ where the experiment is sampling values of $\vmin$ up to exactly this cross over. For masses approaching this critical value from below it becomes increasingly difficult to detect the stream since the SHM+S1 distribution looks more and more like the SHM only distribution but fitted with larger mass. However once $m_\chi$ exceeds the critical point, the experiment is now sampling values of $v_{\rm min}$ exclusively \emph{below} the cross over point. For this range of speeds the SHM only distribution cannot be fitted to SHM+S1 data simply at a different mass (this is because the $g^{\rm SHM+str}(\vmin)$ always lies below $g^{\rm SHM}(\vmin)$ over this range). The end result in terms of the discovery limits is the sharp peaks seen in Fig.~\ref{fig:streamDL_xenon}. It turns out that this effect can be alleviated somewhat with target complementarity as we will discuss in the next section.

\subsection{Including the annual modulation of the signal}\label{subsec:anmod}

\begin{figure}[t]
\includegraphics[width=0.49\textwidth]{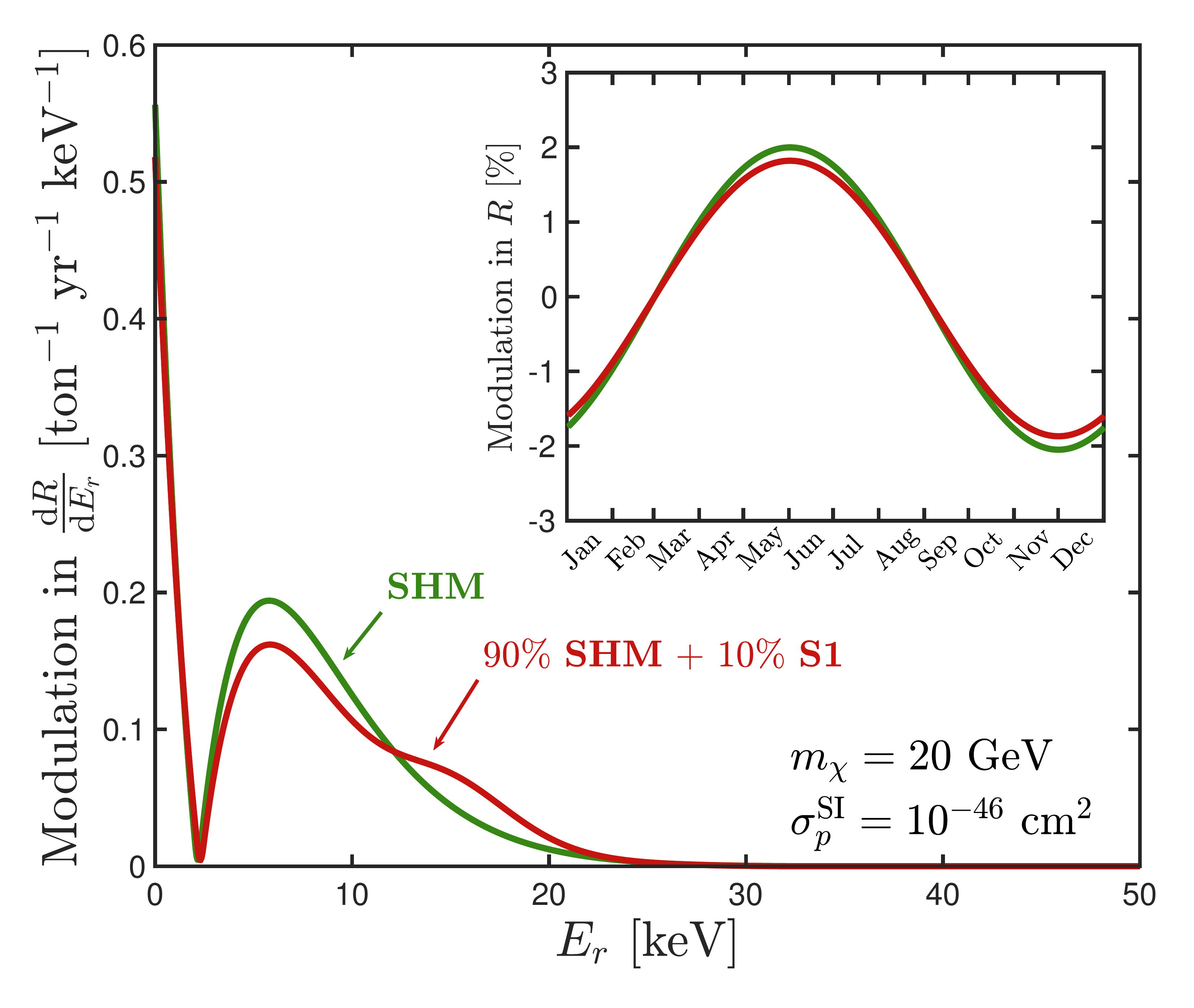}
\caption{{\bf Main:} Annual modulation amplitude in
  $\textrm{d}R/\textrm{d}E_r$ as a function of recoil energy. The green lines show modulation under the SHM only model, the red lines after the inclusion of a 10\% S1 component. {\bf Inset:} The
  modulation amplitude of the total rate $R$ for both models as a
  function of time.}\label{fig:annual-modulation}
\end{figure}

 Unfortunately, through DM--nuclear scattering much of the kinematic information about the S1 stream is lost. 
 To improve the range of parameters over which the S1 stream is detectable, we require additional information.
 One option is to exploit the unique annual modulation of the DM signal.
 
 In Fig.~\ref{fig:annual-modulation} we show the annual
modulation signal for a DM mass of~20 GeV for
both the SHM and SHM+S1 models when $\rho_{\rm str}/\rho_0=0.1$.
 The main panel shows the modulation 
amplitude of the differential recoil rate
$\textrm{d}R/\textrm{d}E_r$ as a function of $E_r$. The
modulation amplitude vanishes when
$v_{\rm min}(E_r)$ sits at the stationary point in the time evolution of $f(\mathbf{v},t)$. 

 In the inset panel, we show the modulation of the total event rate
integrated above $E_r = 1$~keV. While the stream does modulate
in phase with the smooth halo, we see that for this WIMP mass and
threshold it actually slightly decreases the modulation of the total
rate. This is because the stream (as a fraction of $\rho_0$) takes
some low energy recoils and shifts them to higher energies. This can be
 seen in the main panel as an enhancement for recoil energies between 10 and 20 keV. 
At these
higher energies the amplitude is decreasing, meaning that
the overall modulation of the total rate gets suppressed. 

Annual modulation is a useful signature for WIMP discovery. The backgrounds we consider here either do not modulate at all (the laboratory and DSNB backgrounds), or they modulate with an entirely different phase and amplitude (Solar and atmospheric neutrinos). Clearly to gain discrimination power from some new signal the essential feature one needs is for that signal to distinguish the two hypotheses. So while this is certainly true for distinguishing a WIMP from the background, unfortunately it is not the case for distinguishing the SHM from the stream. Because S1 modulates with the same phase and leaves the modulation amplitude mostly unchanged, when it we incorporate time dependence into our stream discovery limit calculation we see essentially no impact at all.

Nevertheless there may be other ways to distinguish the two halo models. Additional information could come from the 
complementarity between multiple experiments that exploit different target nuclei~\cite{Pato:2010zk,Peter:2013aha} 
or from neutrino telescopes searching for DM annihilation in the Sun~\cite{Kavanagh:2014rya,Ibarra:2018yxq}. 
These could both potentially alleviate the degeneracy between the WIMP parameters and the stream density, 
but we leave these questions open for future work. Instead, we next explore what we foresee to be the most powerful piece of extra information; that which can be gleaned in directional WIMP detectors.

\section{Directional WIMP detectors}\label{sec:directional}

There is a strong science case to be made for detecting the directionality of a nuclear recoil signal~\cite{Mayet:2016zxu}. The unique angular signature of a signal with a galactic origin facilitates the discovery of WIMPs with, in principle, fewer events than would be required if only recoil energy information is measured~\cite{Green:2010zm}. Furthermore the signal cannot be mimicked by any known terrestrial~\cite{Leyton:2017tza} or cosmic background~\cite{Mei:2005gm}, including Solar neutrinos~\cite{Grothaus:2014hja, O'Hare:2015mda, OHare:2017rag}. 

Realising the measurement of $\mathcal{O}(1-100)$ keV recoil tracks is challenging. In liquid or solid state detectors recoil tracks are typically nm-sized, whereas in gas they can be on the order of a few~mm. This means that a directional detector requires either a readout method with incredibly high spatial resolution (e.g.\ the X-ray imaging of nuclear emulsions in NEWSdm~\cite{Aleksandrov:2016fyr,Agafonova:2017ajg}) or detection media with very low pressure (e.g.\ gas time projection chambers like DRIFT~\cite{Daw:2011wq,Battat:2014van,Battat:2016xaw}, DMTPC~\cite{Monroe:2011er,Leyton:2016nit}, MIMAC~\cite{Santos:2011kf,Riffard:2013psa} and NEWAGE~\cite{Nakamura:2015iza}). What these methods gain in directional sensitivity they lose in their overall practical size. Hence the discovery power of directional detectors still trails behind the more mature non-directional WIMP detectors. Nevertheless, much progress has been made in the development of sophisticated readout technologies for measuring mm-scale tracks in gas. 

The directional detection community is currently establishing which of these new technologies is most powerful and cost-effective when multiplied over the large readout planes that are needed for chambers that can hold ton-scale target masses in the gas phase. A design for such a detector called CYGNUS has been proposed and a feasibility study is currently being conducted~\cite{CYGNUS}. 
We use this feasibility study as the basis of our analysis for a realistic, future directional detector. 

\subsection{CYGNUS}

The preliminary design study for CYGNUS is for a gaseous time projection chamber with a total active volume between 1000~m$^3$ and 100,000~m$^3$
of $^4$He:SF${_6}$ gas, likely to be in a modular and/or multi-site setup. Although the precise setup is the subject of ongoing refinement, the current suggestion is for the chamber to hold the SF$_6$ gas at 20 torr, or He at 740 torr. At room temperature, 1000 m$^3$ of SF$_6$ at 20 torr and He at 740 torr both have masses of 0.16 tons. Since helium tracks in gas are much longer than fluorine, a much higher pressure mode is possible. For the CYGNUS detector there is also the possibility of a `search mode' experiment with 200 torr SF$_6$. This would have limited directional sensitivity but would increase the experiment's exposure by a factor of~10. 

For a given readout technology with a fixed spatial resolution, the energy threshold is set by the limit below which all directional information on an event is lost. The main effects that reduce the directionality of a track are caused by diffusion of the ionisation cloud as it drifts to the readout plane and `straggling' as the initial recoil scatters off other nuclei. For helium recoils in 740 torr, this limit is found to be around 1~keV whereas for fluorine recoils at 20 torr, it is around 3~keV. We use these values as the respective energy thresholds in our analysis. 

An issue for directional detectors is head-tail recognition, i.e.\ measuring the sign of the direction~$\hat{\textbf{q}}$ associated with each nuclear recoil~\cite{Dujmic:2008zz,Majewski:2009an,Billard:2012bk,Battat:2016xaw}. In principle both the charge deposition and the track topology should give an indicator of the head/tail of an event. In practice however, diffusion limits how well the charge distribution can be used to infer the head or tail, and the topology can be measured less well for shorter tracks. Both of these effects worsen at lower recoil energies so in our detector model, we set an additional threshold below which we can no longer measure the sign of~$\hat{\mathbf{q}}$. 

Various readout technologies are compared in Ref.~\cite{CYGNUS}. For our study, we assume that the readout can perform 3-dimensional track reconstruction with an angular resolution of 30$^\circ$ for fluorine and helium recoils above their respective threshold energies. We assume a head-tail recognition efficiency of~100\% above 10~keV and 50\% (i.e.\ no head-tail recognition) below 10~keV. This performance is realistic for readouts based on pixel grids or orthogonal conducting strips. For a much more comprehensive review of readout technologies for the directional detection of DM, see Ref.~\cite{Battat:2016pap}.

Finally, we make the assumption that CYGNUS has perfect electronic/nuclear recoil discrimination. This is a reasonable assumption since the track topologies of electrons and nuclei in gas are so distinct that even the most rudimentary of readout technologies can achieve very high discrimination power. We include a nuclear recoil background comprised of the same set of Solar, DSNB and atmospheric neutrinos, as well as an isotropic and flat laboratory background. Details on the analytic calculation of the directionality of the neutrino background are described in Ref.~\cite{O'Hare:2015mda}.

\subsection{Distinctive features in the angular recoil rate}

\begin{figure*}[t]
\centering
\includegraphics[width=\textwidth,trim={0cm 3.5cm 2cm 0cm},clip]{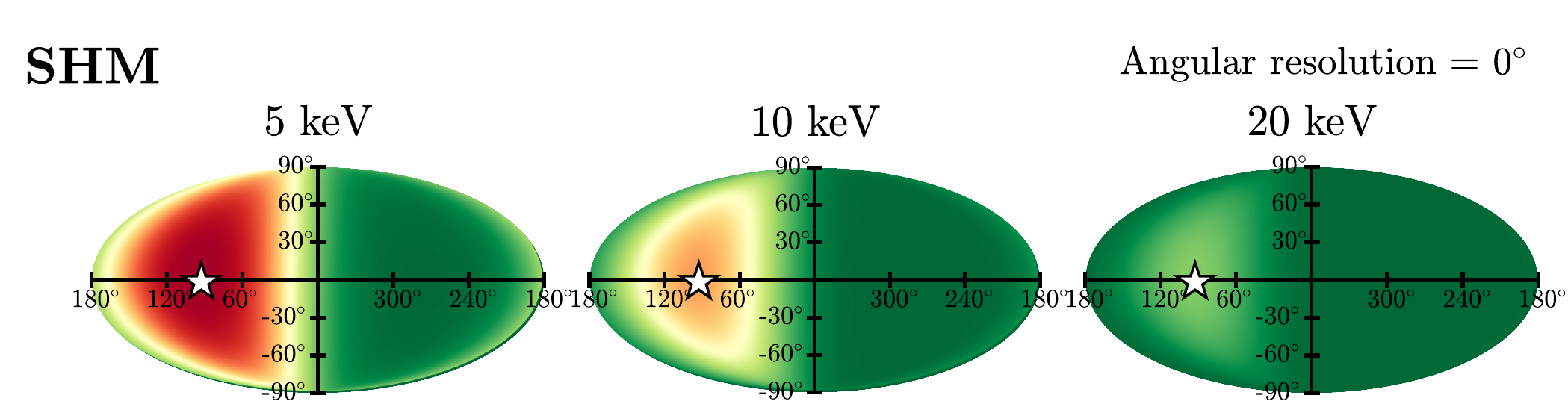}
\includegraphics[width=\textwidth,trim={0cm 3.5cm 2cm 0cm},clip]{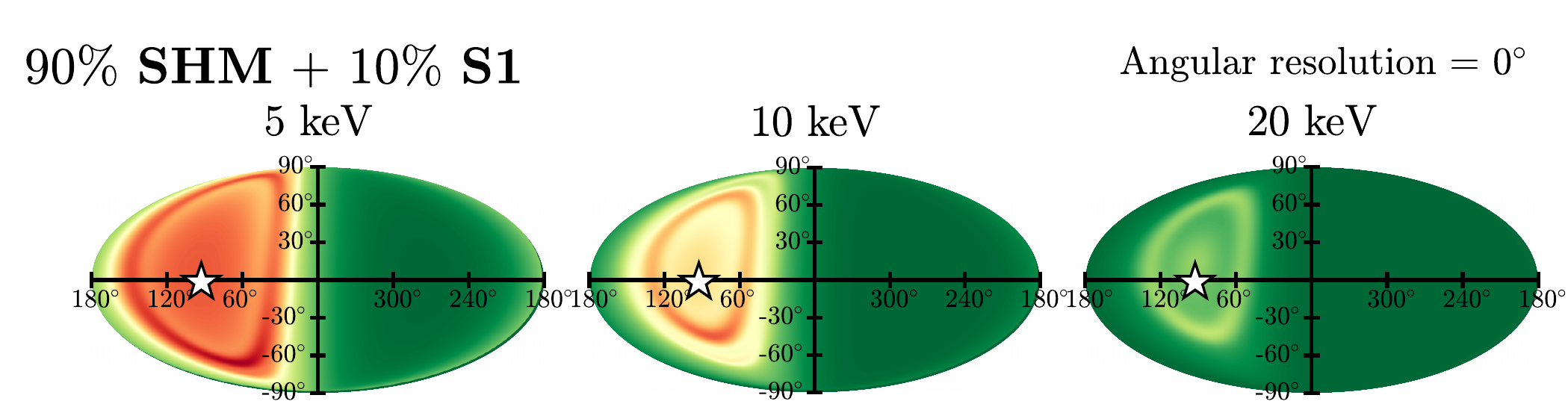}
\includegraphics[width=\textwidth,trim={0cm 0cm 2cm 0cm},clip]{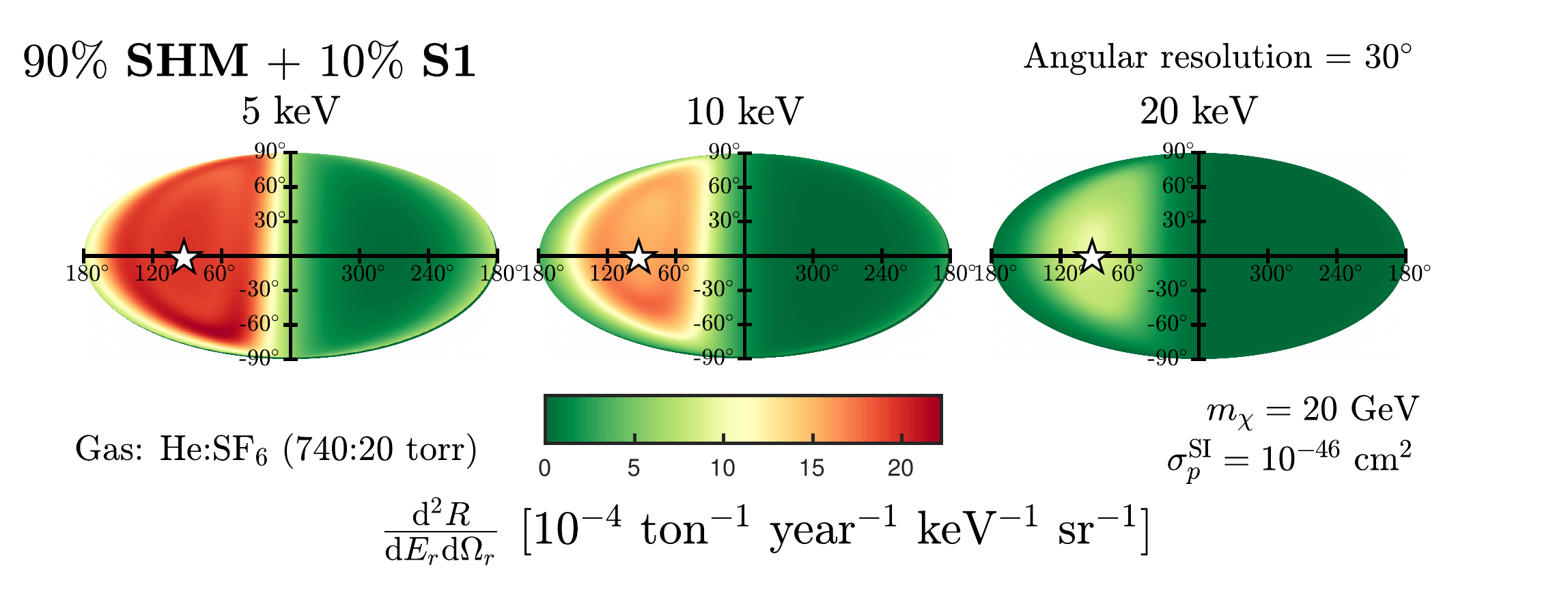}
\caption{Mollweide projection in galactic coordinates of the value of
  the double differential angular recoil rate as a function of the
  inverse of the recoil direction $-\qhat$ and as a function of
  energy (5, 10 and 20 keV from left to right). We assume a 20 GeV WIMP
  and include the angular recoil spectra from both He and SF$_6$
  gas with pressures of 740 and 20 torr
  respectively. The upper three panels show the angular distribution of the SHM model, the middle panels show the distribution after the inclusion of a 10\% contribution from S1, while the lower three panels show the same distribution as the middle three but after a smearing by an angular resolution of 30$^\circ$. In each panel we indicate the
  direction of $\vlab$ with a white star.
  }\label{fig:directional}
\end{figure*}

As we have just discussed, directional WIMP detectors such as CYGNUS measure
the direction~$\hat{\textbf{q}}$ associated with each nuclear recoil in addition to the nuclear recoil energy~$E_r$.
In analogy with Eq.~(\ref{eq:finaleventrate}), the double
differential event rate as a function of recoil energy, recoil direction and time~is
\begin{equation}\label{eq:finaleventrate-directional}
 \frac{\textrm{d}^2R(t)}{\textrm{d}E_r\textrm{d}\Omega_r} = \frac{\rho_0\,\sigma^{\rm SI,SD}_p}{4\pi\mu_{\chi p}^2 m_\chi}  \mathcal{C}_{\textrm{SI,SD}} F^2_{\rm SI,SD}(E_r) \, \hat{f}(\vmin,\hat{\textbf{q}},t) \, .
\end{equation}
This formula is similar to the non-directional rate, except we have
picked up a factor of $1/2\pi$ and require
$\hat{f}(\vmin,\hat{\textbf{q}},t)$ instead of $g(\vmin,t)$. This is
the `Radon transform' of the velocity
distribution~\cite{Gondolo:2002np,Radon},
\begin{equation}
 \hat{f}(\vmin,\hat{\textbf{q}},t) = \int \delta\left(\textbf{v} \cdot \hat{\textbf{q}} - \vmin\right) f(\textbf{v},t)\, \textrm{d}^3 \textbf{v}\, .
\end{equation}
For directional detectors, $\hat{f}(\vmin,\hat{\textbf{q}},t)$ contains all of the dependence
on the DM velocity distribution so this is where the difference between the smooth, isotropic SHM halo
and SHM+S1 model enters.

Since CYGNUS will have the spin-carrying $^{19}$F as a target nucleus, we also allow for the possibility of spin-dependent (SD) scattering in addition to spin-independent (SI) scattering considered in Sec.~\ref{sec:xenon}.
Fluorine carries a nuclear spin $J = 1/2$ and has a relatively high proton spin expectation value of $\langle S_p \rangle  = 0.42$~\cite{Cannoni:2012jq}.
This means that directional detectors, which often use $^{19}$F targets, are well suited to set competitive constraints on $\sigma_p^{\mathrm{SD}}$, the SD WIMP-proton scattering cross section. 

The directional differential scattering rate in Eq.~\eqref{eq:finaleventrate-directional} is valid for DM-nucleus scattering with a single nuclear species. 
For detectors such as CYGNUS that contain multiple target nuclei, the total rate is obtained by summing Eq.~\eqref{eq:finaleventrate-directional}
over all target nuclei weighted by their fractional abundances within the detector.
For SI scattering, $F_{\mathrm{SI}}(E_r)$ is parameterised by the Helm nuclear form factor 
and the value of $\mathcal{C}_{\rm SI}$ given in Eq.~\eqref{eq:CSI}.
The nuclear enhancement factor in the SD case is
\begin{equation}
\mathcal{C}_{\rm SD} = \frac{4}{3} \frac{J+1}{J} \bigg| \langle S_p \rangle + \left(\frac{a_n}{a_p} \right)\langle S_n \rangle \bigg|^2\, .
\end{equation}
For $^{19}$F, $\langle S_n \rangle$ is negligible~\cite{Cannoni:2012jq} so in our analysis, we assume a proton-only coupling scenario where $a_p = 1$, $a_n = 0$. We make use of the shell model calculations of Ref.~\cite{Divari2000} for fluorine recoils. In principle we expect some recoils from sulphur as well. For SI interactions the total rate for $^{32}$S is only a factor 2 smaller than fluorine when we account for the $A^2$ enhancement and the atomic ratio of SF$_6$. However since sulphur tracks are shorter in general we would need a lower pressure and higher energy threshold to achieve decent background rejection.

\begin{figure*}[t]
\includegraphics[width=0.49\textwidth]{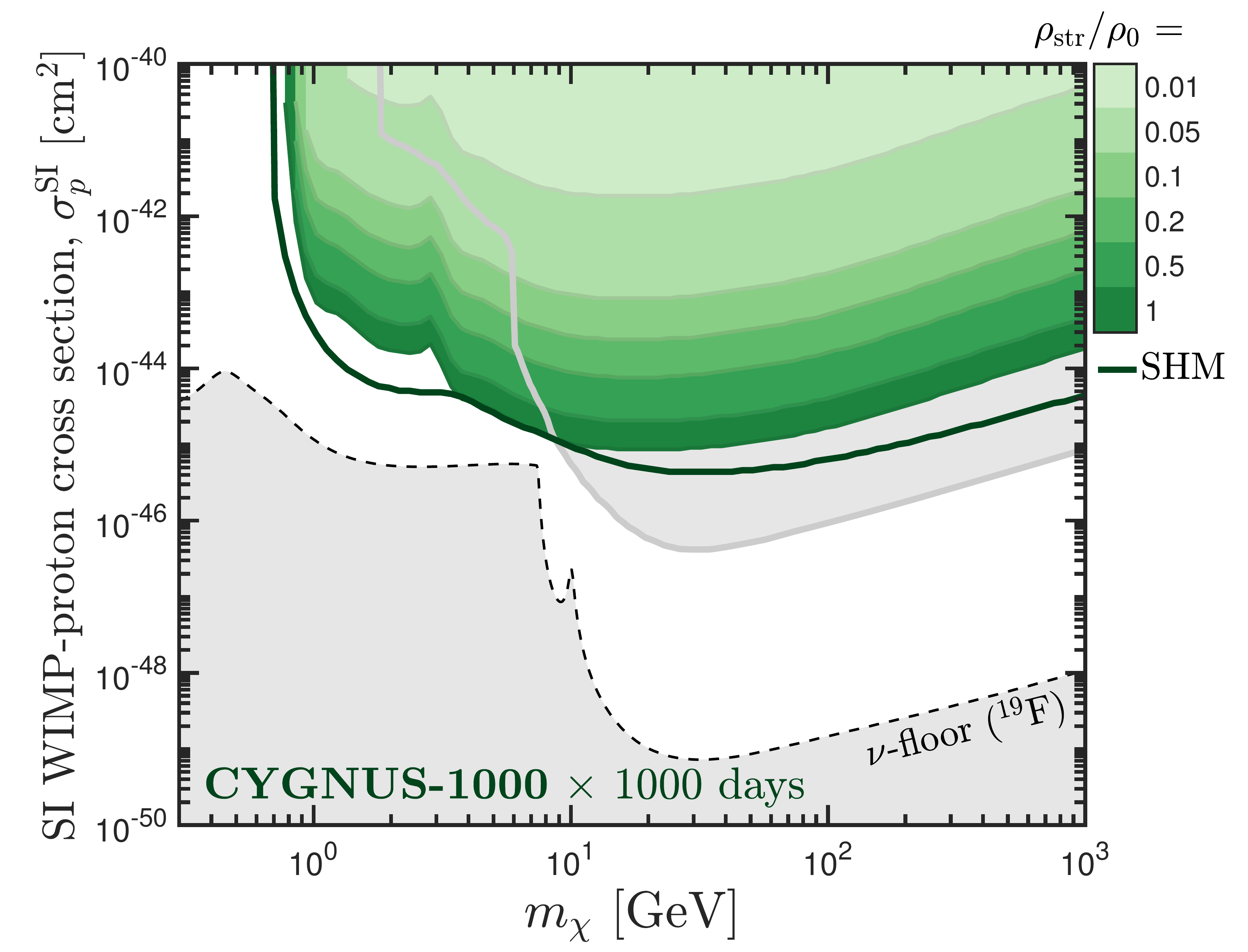}
\includegraphics[width=0.49\textwidth]{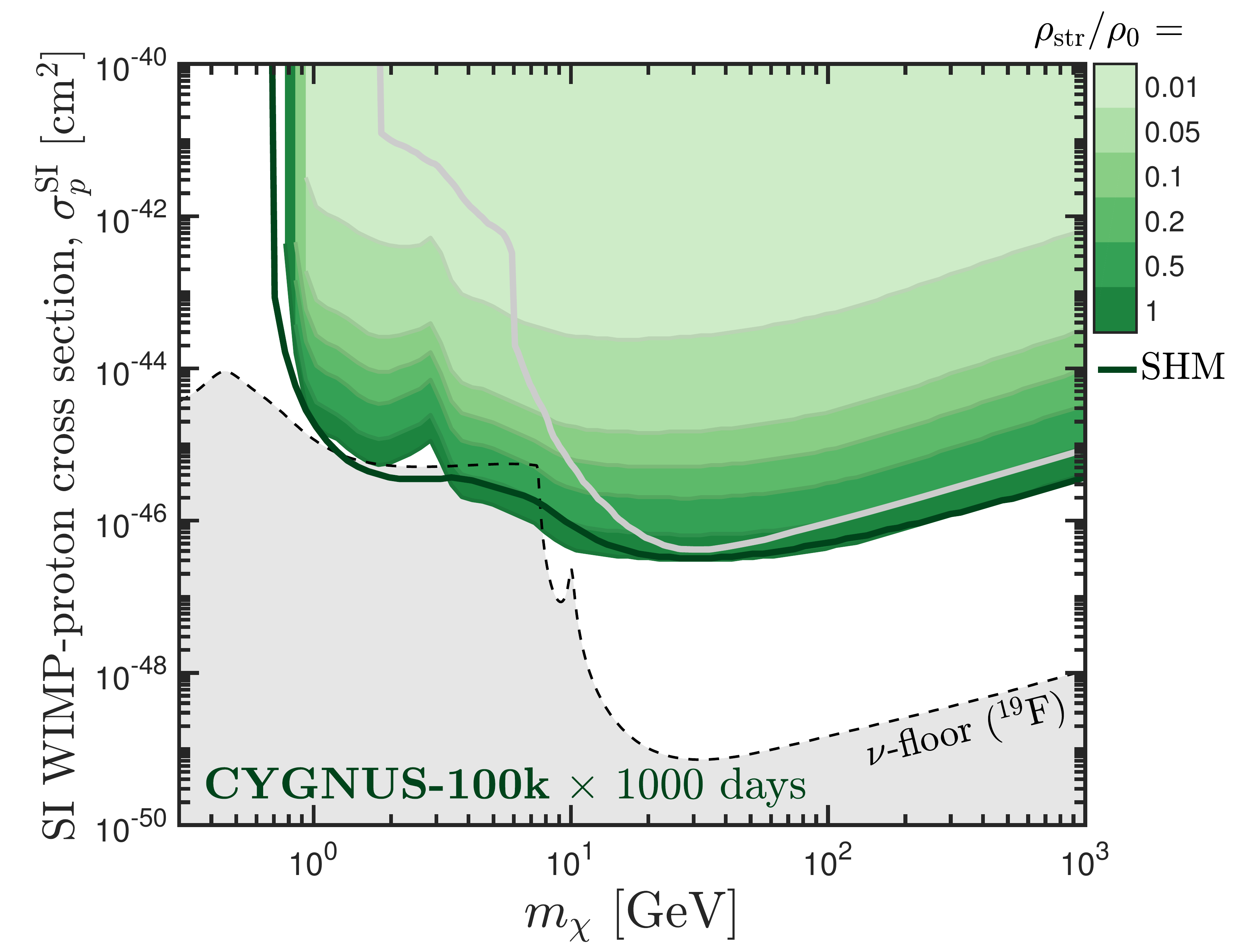}
\includegraphics[width=0.49\textwidth]{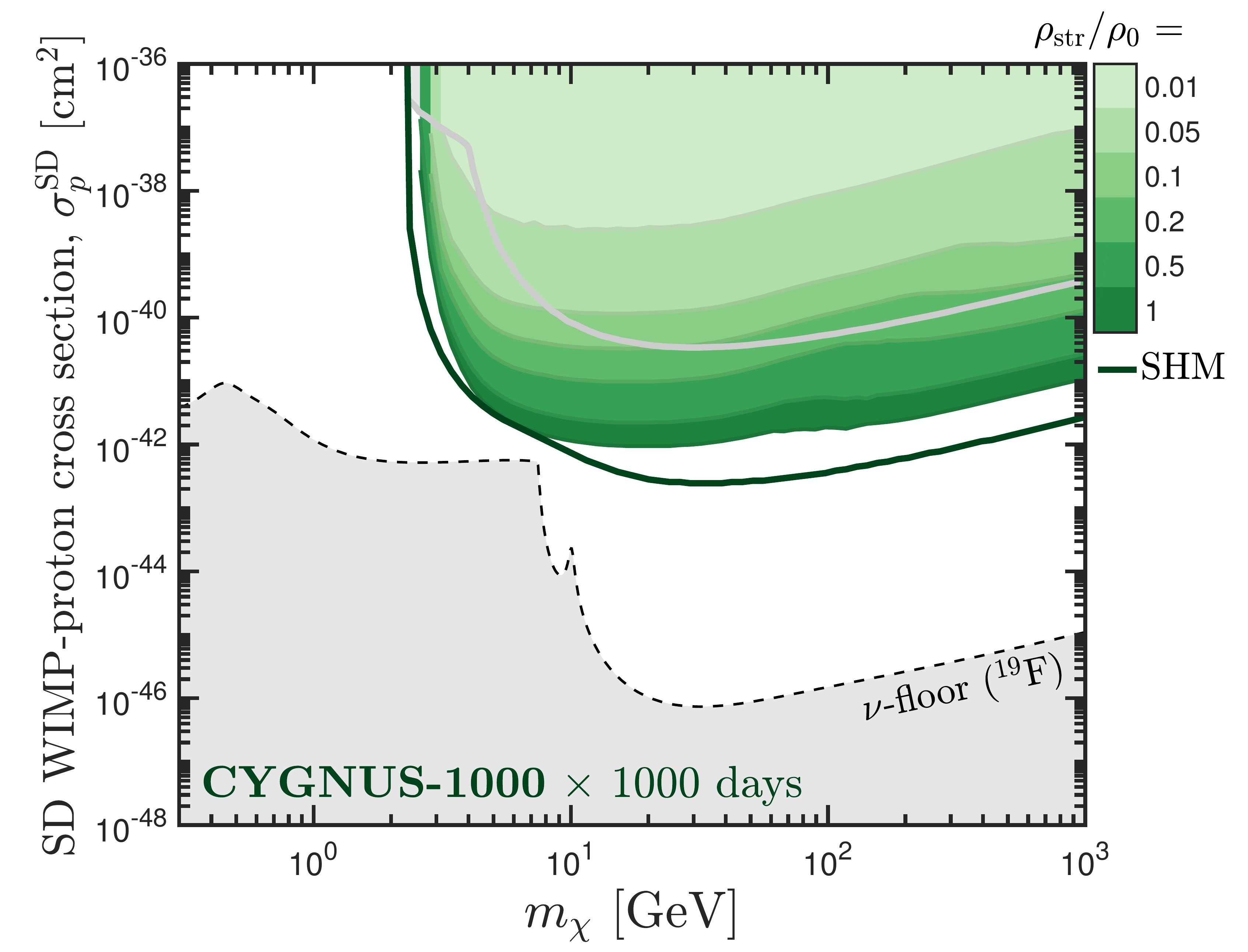}
\includegraphics[width=0.49\textwidth]{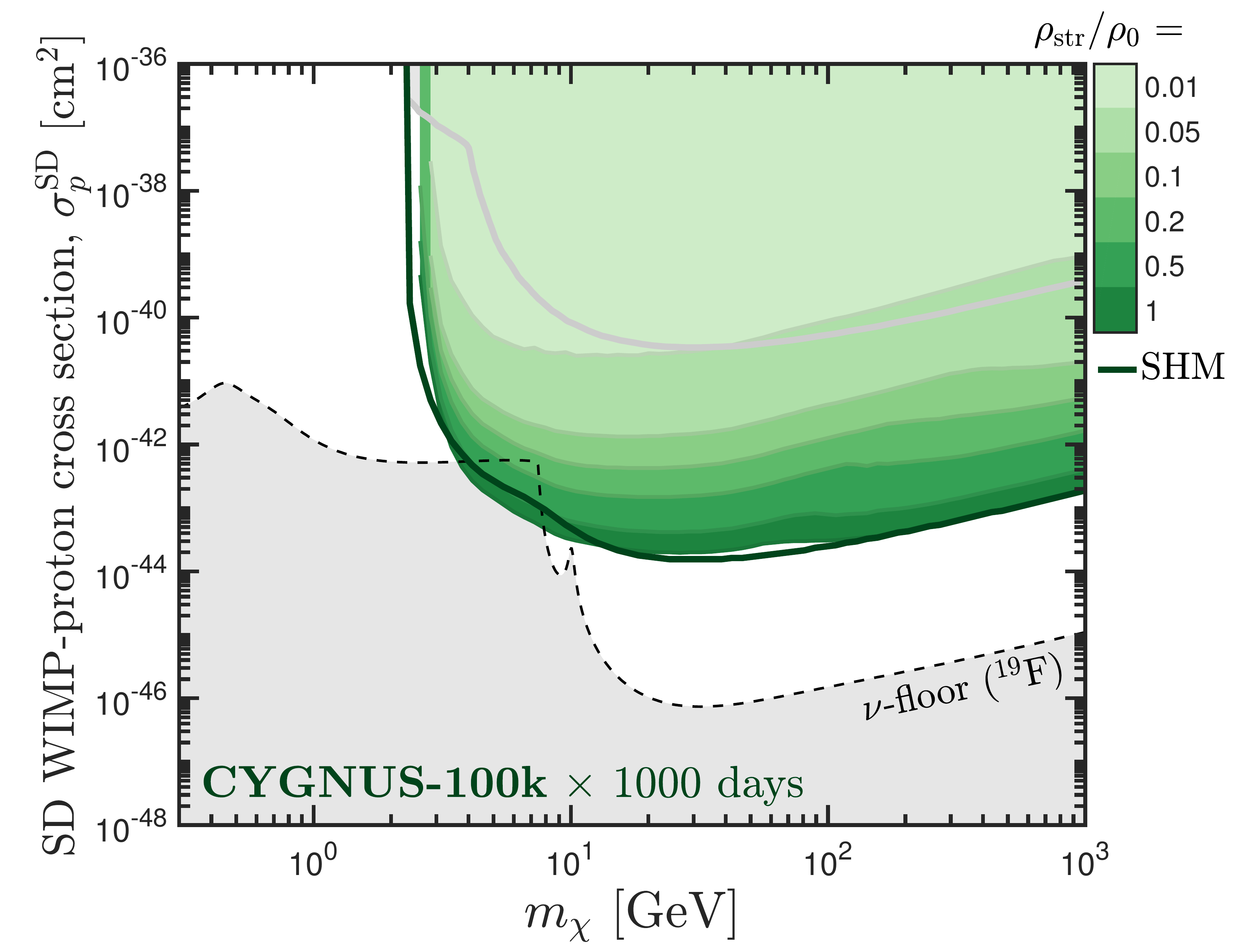}
\caption{SI (top) and SD (bottom) stream discovery limits as in Fig.~\ref{fig:streamDL_xenon} but for the CYGNUS-1000 (left) and CYGNUS-100k (right) experiments. In this case we show the neutrino floor for a $^{19}$F target. In the SI case we can exclude lower masses due to the presence of helium recoils. In the SD case the S1 stream can be distinguished from the smooth halo at much higher masses because the SD WIMP-proton cross section is less well constrained.}\label{fig:streamDL_cygnus}
\end{figure*}

We show angular recoil distributions at three individual recoil energies in Fig.~\ref{fig:directional}. 
A comparison of the upper and middle panels, for the SHM and SHM+S1 models respectively, 
shows that ring features that decrease in angular radius with energy are present in the SHM+S1 model.
The differential rate shown in Fig.~\ref{fig:directional} assumes SI  scattering, so at a given recoil energy, 
there are two rings owing to nuclear recoils from helium and fluorine.  The two rings are most clearly seen in the middle panels, where the angular resolution has been ignored,
but even when a realistic angular resolution is applied as in the bottom panels, the ring features are still somewhat present.

A similar feature does appear in the recoil distribution of the bulk halo at low recoil energies and for heavy DM masses, as was studied in Ref.~\cite{Bozorgnia:2011vc,Mayet:2016zxu}. However the feature appears much more prominently here because the stream's velocity distribution $f_{\mathrm{str}}(\mathbf{v},t)$ is tightly focused around one direction. For a given incoming speed there is an exact kinematic relationship between a single recoil energy and scattering angle. The angular radius of the ring (i.e. the angle between $\vlab-\vstr$ and a point on the ring) at a given energy $E_r$ is given by,
\begin{equation}\label{eq:scatteringangle}
\cos\theta_{\rm str} = \frac{1}{|\vlab - \vstr|}\sqrt{\frac{m_N E_r}{2 \mu_{\chi N}^2}} \, .
\end{equation}
The angular width of the ring for a given recoil energy is due to the stream dispersion and increases slightly with energy, $\Delta \cos{\theta_{\rm str}} \simeq \sigma_{\rm str}\cos{\theta_{\rm str}}/|\vlab - \vstr|$. The center of the ring is slightly shifted away from the direction of $\vlab$ (indicated by the star in Fig.~\ref{fig:directional}), reflecting the slight inclination of S1 away from the plane of the disk.

Streams generally increase the degree of anisotropy in the angular recoil spectrum because they are more focused around a particular direction. 
The direction of S1 opposes galactic rotation so it particularly adds to the anisotropy because its recoils are in the same direction as the dipole from the SHM.
In addition, S1 adds recoils at higher energies, which from Eq.~\eqref{eq:scatteringangle} must have smaller scattering angles, further increasing the anisotropy.
For the specific case of a 10\% S1 component, the forward-backward anisotropy of the signal above the energy threshold increases by an amount between 25\% and 20\% for masses between 1 GeV and 1000~GeV respectively. Owing to the increased anisotropy of the signal, we anticipate that S1 will aid the discovery reach of a directional detector. 

The results in Fig.~\ref{fig:directional} assume an SI scattering cross section of $\sigma^{\rm SI}_p = 10^{-46}$ cm$^2$. 
A similar scattering rate would be obtained for SD scattering with $\sigma_p^{\rm SD} = 5\times10^{-44}$~cm$^2$.
However, an important difference is that the SD case contains a single ring since scattering only occurs with fluorine. When only considering fluorine the anisotropy increases with a 10\% S1 component by 30 (20)\% for a 1 (1000) GeV mass. This value is slightly larger relative to the SI case mainly because we have a lower energy threshold for helium recoils.

\subsection{S1 discovery limits}

From Fig.~\ref{fig:directional}, we anticipate already that directional detectors should be more powerful for detecting streams since the SHM and SHM+S1 models are visibly different in their angular spectra, wheres they only showed small differences in their recoil energy spectra (cf.\ Fig.~\ref{fig:xenon-rate}).
The consequences of general streams in directional WIMP detectors were studied in detail in Ref.~\cite{O'Hare:2014oxa}.
The analysis of the S1 stream is greatly simplified since we have a known direction in which to look. 
This means we can apply the statistical test described in Sec.~\ref{sec:streamDLs} and make a clear comparison with xenon detectors. 

Figure~\ref{fig:streamDL_cygnus} shows the S1 stream discovery limits for CYGNUS-1000 m$^3$ (left panels) and CYGNUS-100k m$^3$ (right panels)
for a range of values of $\rho_{\rm{str}}/\rho_0$.
The upper panels show discovery limits for SI scattering, while the lower panels show results for SD scattering.
The solid dark green lines show the WIMP discovery limits for the SHM model. 
The upper solid grey lines and upper grey shaded regions in Fig.~\ref{fig:streamDL_cygnus} show
the existing exclusion limits on the scattering cross section.
For the SI DM--proton cross section, the exclusion limits are the same as those in Fig.~\ref{fig:streamDL_xenon}
while for SD scattering, the limits come from PICO-60~\cite{Amole:2015pla} and PICASSO~\cite{Archambault:2012pm}.
The lower shaded region is the neutrino floor for a fluorine target. We do not show the helium neutrino floor for clarity since it is extremely similar to the floor for fluorine.

Comparing with Fig.~\ref{fig:streamDL_xenon}, the most notable difference in Fig.~\ref{fig:streamDL_cygnus} 
is the absence of a rapid rise in the stream discovery limit
at $\sim50$~GeV. For xenon detectors, the SHM and SHM+S1 recoil energy distributions 
became largely indistinguishable above this mass, meaning that the stream could not be measured.
When directional detectors add the additional information about the nuclear recoil direction~$\hat{\textbf{q}}$,
the discovery limits much more closely follow the shape of the SHM sensitivity curve.

The SI discovery limits in Fig.~\ref{fig:streamDL_cygnus} extend to much lower values of the DM mass than Fig.~\ref{fig:streamDL_xenon}.
This is a consequence of the light $^4$He nucleus within CYGNUS and low threshold (1 keV),
enabling the detection of lighter DM particles. 
We therefore see that there is good complementarity between CYGNUS and the xenon
detectors for SI scattering: while xenon detectors can discover the S1 stream in the range 
between~5 and~25 GeV, CYNGUS extends the sensitivity down to ~0.8 GeV.
We also see that CYGNUS can probe a much lower values of $\rho_{\rm{str}}/\rho_0$ than a xenon detector,
largely because the exclusions limits on $\sigma^{\rm SI}_p$ for DM below 5~GeV are substantially weaker 
than the limits at higher masses. 
The stringent exclusion limits at high mass mean that if the DM interacts with a SI interaction,
both directional and non-directional detectors will struggle to discover the S1 stream
for DM with a mass greater than approximately 25~GeV.

The lower panels in Fig.~\ref{fig:streamDL_cygnus} show the discovery limits for the SD WIMP-proton cross section $\sigma^{\rm SD}_p$.
There are two major changes with respect to the SI case. The first change is that the discovery limits do not extend below 3~GeV.
This is because SD interactions do not cause helium to recoil, which was the driver of the SI sensitivity at lower mass.
The second change is that the discovery limits at high masses lie below the current exclusion limits.
This is because $\sigma^{\rm SD}_p$ is less constrained while CYGNUS contains a large number of $^{19}$F nuclei
that are especially sensitive to $\sigma^{\rm SD}_p$. 
If the DM interaction with nuclei is SD, the added information from directional detectors provides
essential information that allows the SHM+S1 model to be distinguished from the SHM model
across a much wider range of WIMP masses.

Another slight difference between the SD and SI discovery limits can be seen in the shape of the limits at high mass. 
In the latter it appears that as we increase $\rho_{\rm str}/\rho_0$ there are a series of bumps appearing at decreasing masses. 
We identify these to be a manifestation of the same effect that was giving rise to the sharp peaks in our xenon limits in Fig.~\ref{fig:streamDL_xenon}. 
Previously this was due to the cross over in the shapes of $g(\vmin)$ for the SHM and SHM+S1 models. 
Here, this effect is not as severe because it much harder to mimic the unique 3-d shape of $\hat{f}^{\rm SHM+S1}(\vmin,\qhat)$ without a stream component. 
The reason there seem to be no bumps to speak of in the high mass SI limits is because we are gaining discrimination power not only from directionality, 
but from the complementarity between helium and fluorine. Since scattering from helium does not contribute in the SD case, the bumps return.

We see from the right panels of Fig.~\ref{fig:streamDL_cygnus}
that if the S1 density is very large and for the large exposure assumed in the CYGNUS-100k setup, 
the parameter space below the $^{19}$F neutrino floor (at $\sim 6$~GeV) can be explored.
The degree to which the neutrino background can be subtracted is controlled mostly 
by the angular separation between the WIMP and Solar neutrino dipoles~\cite{O'Hare:2015mda,OHare:2017rag}. 
If CYGNUS-100k could improve the angular resolution below 30$^\circ$ and improve
the head-tail recognition threshold, more of the parameter space below the neutrino floor
could be explored.\footnote{A subtlety in gaining the best discrimination between WIMP 
and neutrino signals in directional detectors is that timing information must be included.
See Ref.~\cite{O'Hare:2015mda} for further details.
}

Directional detector technology is still rapidly evolving and 
we anticipate that if a higher threshold but much larger scale alternative to CYGNUS were feasible,
then competitive limits on SI interactions at high masses may also be achievable.
This could be a possibility in, for instance, a nuclear emulsions detector like NEWSdm~\cite{Aleksandrov:2016fyr,Agafonova:2017ajg},
or if columnar recombination~\cite{Nygren:2013nda,Li:2015zga} can be reliably used as a directional signal 
in liquid xenon~\cite{Nakajima:2015dva,Nakajima:2015cva,Oliveira:2015una,Mohlabeng:2015efa,Nakamura:2018xvy} 
or argon~\cite{Cao:2014gns,Cadeddu:2016mac,Cadeddu:2017ebu}.
So far only 2-dimensional tracks with no head-tail information have been shown to be measurable in nuclear emulsions, 
and only 1-dimensional tracks are theoretically possible with columnar recombination.
Incomplete recoil vectors reduce the directional sensitivity relative to CYGNUS
but this may be balanced in an experiment such as NEWSdm, which is comprised of a mixture of 8 different target nuclei,
potentially giving a high level of target complementarity.
We leave the detailed exploration of the specific impact of the technical restrictions 
on directional detectors --- and there are many --- for future work.

\section{Axion haloscope}\label{sec:axions}
The motivation for axions~\cite{Weinberg:1977ma,Wilczek:1977pj} originates in the dynamical solution of Peccei and Quinn~\cite{Peccei:1977hh}
to the strong-CP problem of quantum chromodynamics (QCD) (see e.g. Ref.~\cite{Kim:2008hd} for a review). 
It has long been known~\cite{Preskill:1982cy,Abbott:1982af,Dine:1982ah} that axions also meet the required properties and cosmological abundance of cold DM in a wide range of production scenarios in the early Universe (see e.g. Ref.~\cite{Marsh:2015xka} for an overview as well as Refs.~\cite{Graham:2018jyp,Klaer:2017ond,Gorghetto:2018myk,Nelson:2018via,Guth:2018hsa,Visinelli:2018wza,Vaquero:2018tib} for recent developments).
Although still considered to be the second most popular class of candidate for DM behind WIMPs,
the axion is accumulating interest from many corners of the community. 
Much of the interest today is driven by the many diverse methods that can
be employed to search for axions in astrophysics, cosmology and
 in laboratory-based experiments. 
 Below, we briefly review three types of axion `haloscopes' that have the
 most promising chances of detecting axion DM.
For a more detailed discussion of experimental searches 
 for axions and axion-like particles, we refer the reader to Ref.~\cite{Irastorza:2018dyq}.

The most common technique to detect axion DM directly on Earth 
is to amplify the signals produced due to the mixing between 
electromagnetic fields and the oscillating local axion field. 
The axion, $a$, is coupled to quantum electrodynamics (QED) through the interaction,
\begin{equation}\label{eq:axionlagrangian}
\mathcal{L} = -\frac{g_{a\gamma}}{4} \, a\, F_{\mu \nu} \tilde{F}^{\mu \nu} \, 
\end{equation}
where $g_{a\gamma}$ is the axion-photon coupling,  
$F_{\mu \nu}$ is the Maxwell tensor 
and $\tilde{F}_{\mu \nu}$ is its dual.

For QCD axion models (i.e.\ those that solve the strong CP problem) 
there is a prescribed linear relationship between the axion mass~$m_a$ 
and the axion-photon coupling~\cite{Kaplan:1985dv,Srednicki:1985xd}:
\begin{equation}\label{eq:axioncoupling}
\frac{g_{a\gamma}}{{\rm GeV}^{-1}} = 2.0\times10^{-16}\, C_{a\gamma}\, \frac{m_a}{\mu {\rm eV}} \, ,
\end{equation} 
where $C_{a\gamma}$ is an $\mathcal{O}(1)$ model dependent constant. 
The most commonly quoted benchmark QCD axions are the eponymous 
`KSVZ'~\cite{Kim:1979if,Shifman:1979if} and `DFSZ'~\cite{Dine:1981rt,Zhitnitsky:1980tq} models with $C_{a\gamma} = -1.92$ and $C_{a\gamma} = 0.75$ respectively~\cite{diCortona:2015ldu}. 

More generally, in many extensions of the Standard Model of particle physics,
additional axion fields are predicted, e.g. Refs.~\cite{Svrcek:2006yi,Arvanitaki:2009fg,Jaeckel:2010ni,Cicoli:2012sz,Nelson:1993nf,Chikashige:1980ui}.
This has motivated the concept of axion-like particles (ALPs),
in which the relation Eq.~\eqref{eq:axioncoupling} is not enforced.
Searches for axions therefore probe $g_{a\gamma}$ over many orders
of magnitude to cover a wide range of models.

In a strong magnetic field an axion or ALP will convert all of its kinetic energy into a photon, which has an energy
$\omega = m_a(1+v^2/2)$.\footnote{
The notion of a single axion converting to a photon is rather simplistic. 
DM axions behave locally like a classical field. It is the 
oscillation of the field $a(x,t)$ inside an experiment that is observed.
The axion manifests in time-stream data as a superposition of modes given by the astrophysical distribution $f(\mathbf{v})$.}
Even with the strongest magnetic fields available, the photon flux is very small. 
The signal can however be amplified and 
the most common strategy is to match the axion mass to a resonance condition. 
The best known example is the resonant mode of a cavity.

Resonant cavities are the ideal approach for an axion mass in the range 1--40~$\mu$eV. The main feature of the resonant cavity haloscope is the ability to tune the resonant modes over a wide range of frequencies so that the experiment can set limits over the same range of axion masses. Progress with this haloscope design has long been driven by ADMX~\cite{Asztalos:2009yp,Du:2018uak}. 
Recently though several other groups have adopted this method, 
making the relevant modifications to adapt it to other frequencies, e.g.\ HAYSTAC~\cite{Brubaker:2016ktl,Rapidis:2017ytq,Brubaker:2017rna,Zhong:2017fot,Brubaker:2018ebj}, CULTASK~\cite{Chung:2016ysi,Lee:2017mff,Chung:2017ibl}, 
Orpheus~\cite{Rybka:2014cya}, 
ORGAN~\cite{McAllister:2017lkb,McAllister:2017ern}, 
KLASH~\cite{Alesini:2017ifp} and RADES~\cite{Melcon:2018dba}. 

The search towards much higher or lower values of the axion mass requires radically different designs because of several technical restrictions. 
Principally, at higher frequencies resonators require increasingly small volumes and thus suffer decreasing available signal power. 
The dielectric haloscope MADMAX~\cite{TheMADMAXWorkingGroup:2016hpc,MADMAXinterestGroup:2017bgn,Millar:2016cjp}
 aims to circumvent this problem by abandoning the idea of a resonant volume. 
 A dielectric haloscope supplants the high quality factor of a resonator with constructive 
 interference of the axion-induced electric field between a sequence of finely spaced disks, see~Ref.~\cite{Millar:2016cjp}.
 This allows MADMAX to probe axions heavier than 40~$\mu$eV.

For axions lighter than 1~$\mu$eV, the experiment furthest in development is ABRACADABRA~\cite{Kahn:2016aff,Foster:2017hbq,Henning:2018ogd},
though BEAST~\cite{McAllister:2018ndu} and DM-Radio~\cite{Silva-Feaver:2016qhh} are also aiming to test a similar mass window.
ABRACADABRA uses a toroidal magnet to circulate an induced electric current driven by the oscillation of the axion field. 
Subsequently, this circulation will generate a secondary oscillating magnetic field in the centre of the toroid. 
A pickup loop is placed in the centre so that the oscillating magnetic flux can then be inductively coupled to a SQUID magnetometer.
Unlike ADMX, this setup can be sensitive to a large range of frequencies, but at the cost of sensitivity at any one frequency. 
It is possible, however, to create a resonant behaviour with the insertion of a tuned LC circuit between the pickup loop and the SQUID. 
In this configuration the experiment gains a quality factor from the resonance of the circuit. 
This enhances the sensitivity but only in a very narrow range of frequencies, so as with a resonant cavity, it must be designed to scan over $m_a$. 

\subsection{Signal power in an axion haloscope}

Since we are interested in the astrophysical dependence of a haloscope,
 we will frame our discussion about the axion response for an arbitrary experiment. 
 Taking inspiration from Ref.~\cite{Foster:2017hbq}, we write the signal power $P$ in a general way,
 where the astrophysical dependence is explicit.
 For notational convenience we absorb all experimental factors into a haloscope function~$\mathcal{H}(\omega)$.
 We write this as a function of the photon frequency $\omega$:
\begin{equation}
\frac{\textrm{d}P}{\textrm{d}\omega} = \pi \mathcal{H}(\omega) \,g^2_{a\gamma} \,\rho_0 \,\frac{\textrm{d}v}{\textrm{d}\omega} f(v) \, ,
\end{equation}
where $\mathrm{d}v/\mathrm{d}\omega = (m_a v)^{-1}$ and $f(v)$ is the DM {\it{speed}} distribution in the lab frame.
 For ADMX or a similar resonant cavity, we have
\begin{equation}
\mathcal{H}_{\rm cavity}(\omega) = \kappa\, B^2\, V \,C\, \frac{Q}{\omega_0} \,\mathcal{T}(\omega) \, , \label{eq:power}
\end{equation}
where $\kappa \simeq 1/2$ is the coupling efficiency of a cavity, $B~\simeq 8$ T is the magnetic field, 
$V$ is the cavity volume, $Q\sim 10^5$ is the quality factor of the mode and 
$\mathcal{T}(\omega)$ is the Lorentzian transfer function of the mode itself 
(centred at the resonant frequency $\omega_0$, which when scanning will equal the $m_a$ of interest).
The quality factor of a resonant device will in practice be much wider than the axion bandwidth, 
which has an effective $Q \approx 10^6$ (or potentially even higher for an $f(v)$ with substantial cold streams). 
Cavities also have a geometric form factor~$C$ relating the overlap of the electric and magnetic fields. 
ADMX typically use the TM$_{010}$ mode with $C_{010} =0.692$~\cite{Stern:2015kzo}. 

For MADMAX the treatment is not as simple. The spacing of some $N\sim 80$ disks that produce 
the optimum constructive interference over some bandwidth gives a much more complicated 
frequency dependence. This is parameterised by a boost factor $\beta(\omega)$, since 
MADMAX is not a resonator and does not formally have a $Q$. Nevertheless we can write similarly,
\begin{equation}
\mathcal{H}_{\rm dielectric}(\omega) = \kappa \,B^2\, A \,\frac{\beta^2(\omega)}{m^2_a} \, ,
\end{equation}
where $A$ is the area of a disk. Since the boost factor requires a highly involved transfer matrix formalism to compute~\cite{Millar:2016cjp} we simply adopt simple estimates to the sensitivity of MADMAX outlined in its white paper~\cite{MADMAXinterestGroup:2017bgn}. This assumed 80 disks of 1 m$^2$ area with a 10 T magnetic field scanning over masses between 40--400$\mu$eV at an efficiency of $\kappa = 0.8$.

\begin{figure}[t]
\includegraphics[width=0.49\textwidth]{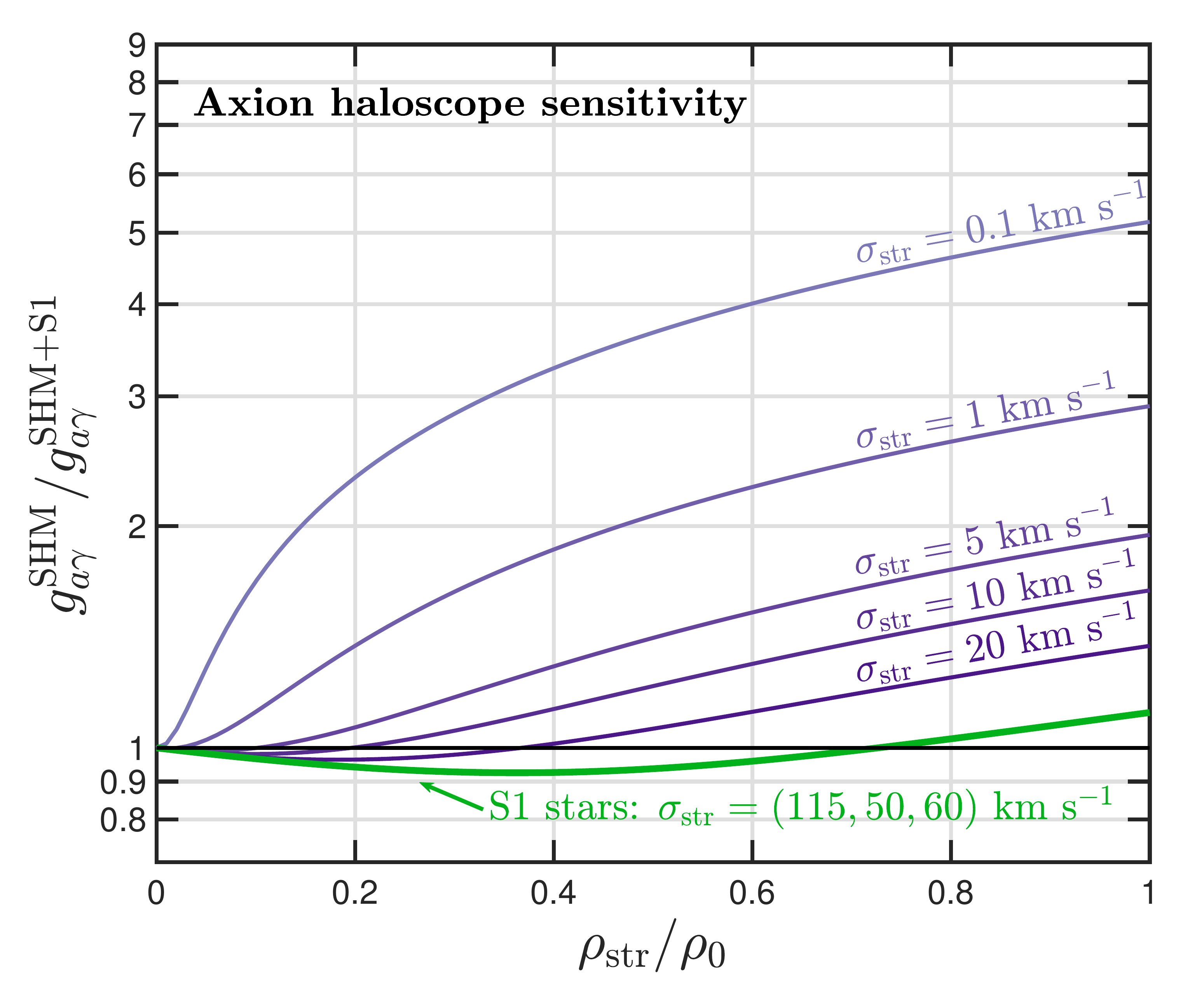}
\caption{Relative change in the sensitivity to the axion-photon coupling for the SHM+S1 model relative to the SHM alone. We make the comparison as a function of the density fraction of S1 and the dispersion of the stream. The dispersion implied by the stellar stream is shown as the green curve; it is equivalent to a 1-dimensional dispersion of $46$~km~s$^{-1}$. Expressed in this way, ratios greater than one have enhanced sensitivity relative to a stream-less halo, while below one the sensitivity is suppressed. Since axions are easiest to detect when their  overall spectrum is sharper, we see that the colder streams give the largest enhancements.}\label{fig:axionstream}
\end{figure}

\begin{figure}[t]
\includegraphics[width=0.49\textwidth]{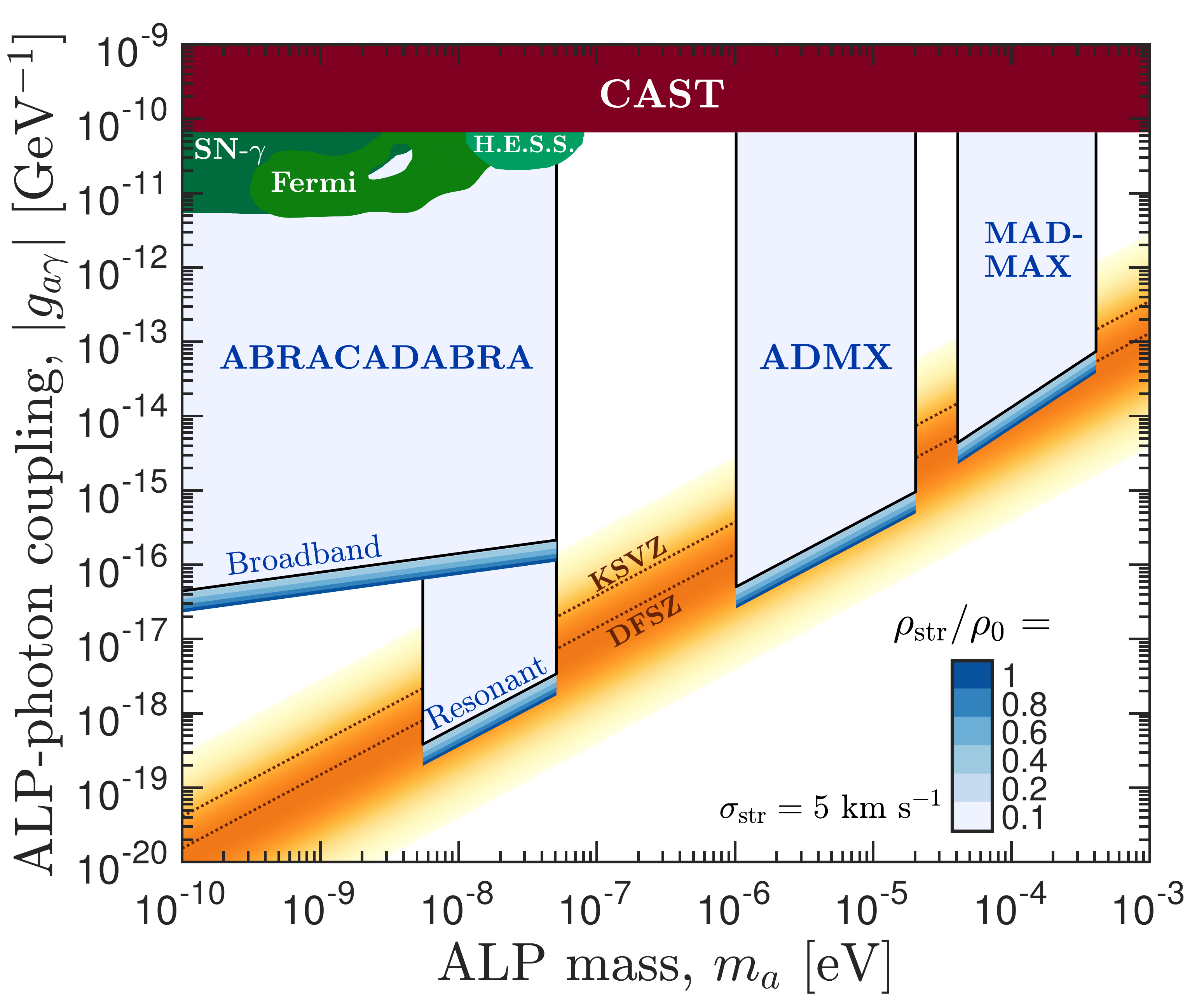}
\caption{The enhancement in the sensitivity of three experiments: ABRACADABRA, ADMX and MADMAX, due to the presence of the S1 stream. In each case we assume projections of these setups to their final QCD sensitive configuration and scan time. For clarity we have shown the result for a very cold analogue of S1 with a stream dispersion of 5 km s$^{-1}$. We also show the already excluded regions of this space from the helioscope CAST~\cite{Anastassopoulos:2017ftl}, and through high energy astrophysical observation (Refs.~\cite{TheFermi-LAT:2016zue,Wouters:2013iya,Payez:2014xsa} labelled Fermi, H.E.S.S. and SN-$\gamma$ respectively). }\label{fig:axionlimits}
\end{figure}

 \begin{figure*}[t]
\includegraphics[width=0.49\textwidth]{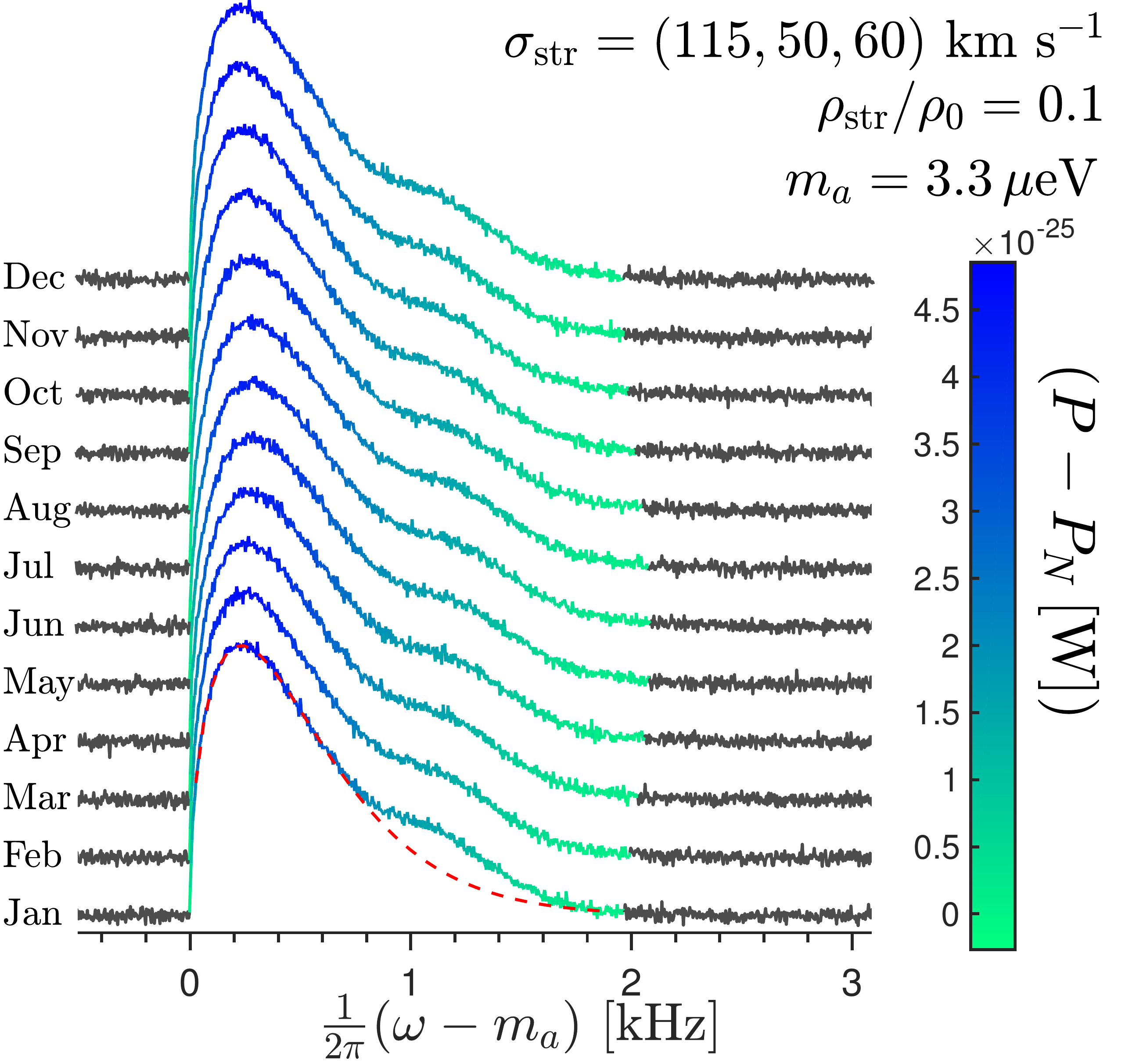}
\includegraphics[width=0.49\textwidth]{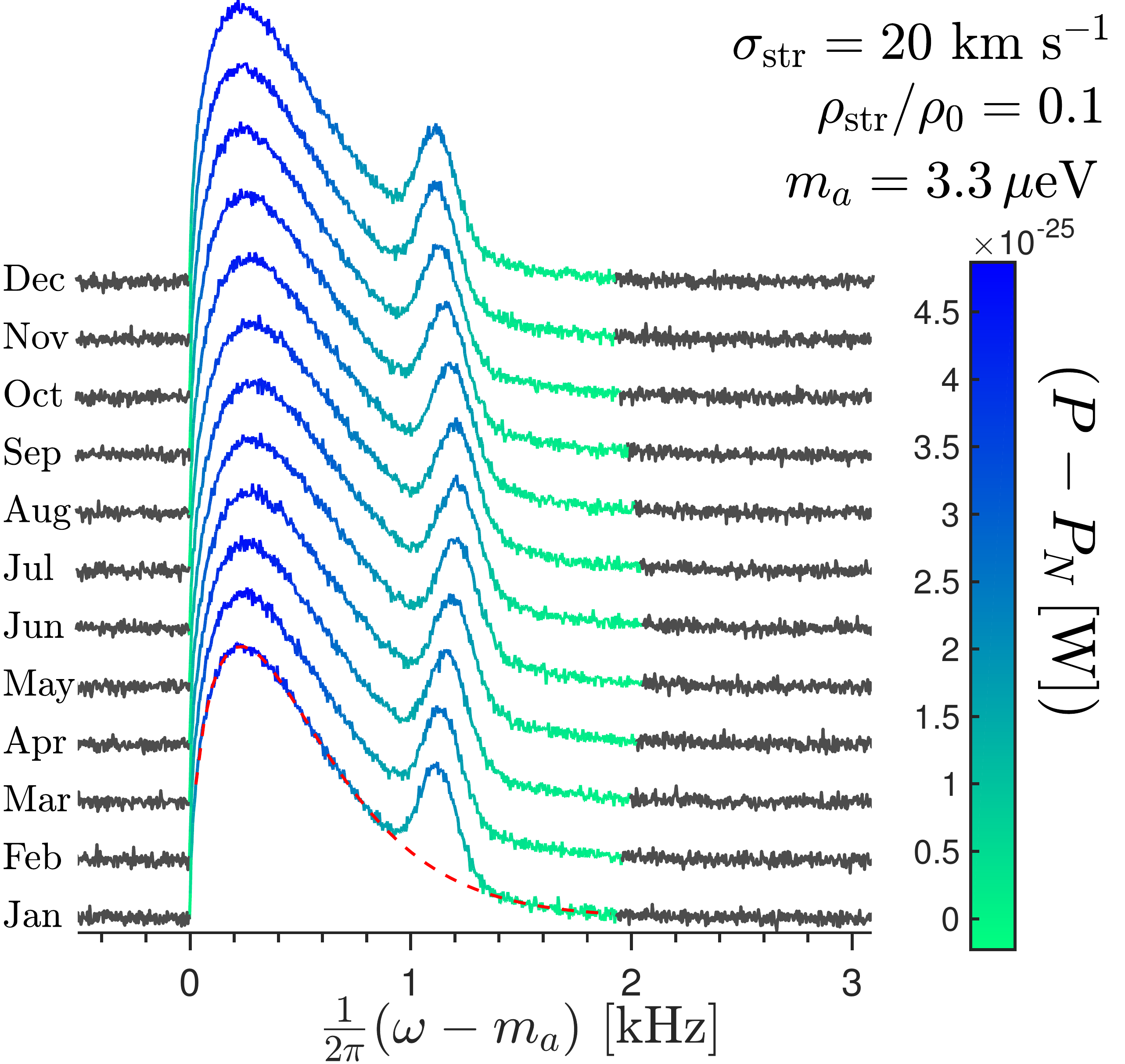}
\caption{Axion power spectra as a function of frequency (shifted by the axion mass $m_a$) and time in months over the year. {\bf Left:} the signal for the SHM+S1 model with the dispersion velocity of the stellar stream, equivalent to a 1-d dispersion of $46$~km~s$^{-1}$ {\bf Right:} the signal for a much colder S1 with a 1-d dispersion of $20$~km~s$^{-1}$. The power has been calculated for an ADMX-like experiment and a DFSZ axion. We assume the power spectra have been obtained for integration times of 0.28 seconds and then averaged over each month. We show Gaussian fluctuations due to thermal white noise for illustrative purposes. For comparison we show the 90\% contribution from the SHM as a red dashed line.}\label{fig:axion_spectrum}
\end{figure*}

Finally, for ABRACADABRA the signal formalism has been detailed thoroughly in Ref.~\cite{Foster:2017hbq}. 
The haloscope functions are, in broadband and resonant configurations,
\begin{equation}
\mathcal{H}_{\rm broad.} = \frac{\alpha^2}{4} \frac{L}{L_p} B^2 V^2 \, ,
\end{equation}
\begin{equation}
\mathcal{H}_{\rm res.}(\omega) = \alpha^2 \frac{L L_i}{(L_p+L_i)^2} B^2 V^2 Q^2 \mathcal{T}(\omega) \, .
\end{equation}
where $\alpha^2 \simeq 0.5$. Here we have now the inductance of the LC circuit $L_i$ and the pickup loop $L_p = \pi R^2/h$ (where $R$ and $h$ are the radius and height of the torus) as well as the inductance of the SQUID labelled $L$. The $V$ here is effectively a geometry factor with dimensions of volume. See Refs.~\cite{Foster:2017hbq, Kahn:2016aff} for further details. 

We emphasise that we have sourced numbers and scanning strategies from analyses that vary in how precisely they might correspond to the final run of one of these experiments. Moreover these projections are not all based on thorough statistical analyses (with the exception of Ref.~\cite{Foster:2017hbq}). However it is still possible to make quite general experiment-independent statements about the detection of the S1 stream.

\subsection{Sensitivity enhancement}

Like in the case for WIMPs, the presence of streams can enhance the prospects for detecting axion DM.
 However, the main difference here is that the sensitivity to axions is controlled by the sharpness of the signal, 
 rather than just its energy distribution. 
 This means that in comparison to the analysis in Secs.~\ref{sec:xenon} and~\ref{sec:directional}, the stream dispersion plays
 a more important role in the observable signal in an axion haloscope.

We parameterise the speed distribution's enhancement or suppression of the sensitivity of a haloscope to DM in the following way:
For an experiment taking time-stream EM signal data, an appropriate likelihood can be constructed from a sum over the power distributed across many frequency bins.
The statistics of the signal+noise is exponential in each bin, 
but for most experiments a stacking procedure of many power spectra can be used to render the distribution Gaussian via the central limit theorem.
Hence the likelihood can be written in terms of a $\chi^2$ sum over frequency bins.
Under the Asimov formalism any test statistic constructed in this way from likelihood ratios can
 be approximated in terms of the integral over the power spectrum squared. 
 Since the power spectrum is proportional to $g^2_{a\gamma}$, the minimum discoverable value will scale as,
 \begin{equation}
g_{a\gamma} \propto \sqrt{\frac{1}{\rho_0}} \left(\int_{m_a}^{\omega_{\rm esc}} \mathrm{d} \omega\, f(\omega)^2 \right)^{-1/4} \, ,
\end{equation}
where the distribution of frequencies is the speed distribution up to a change of variable $f(\omega) = \textrm{d}v/\textrm{d}\omega \, f(v)$. The upper limit is given by $\omega_{\rm esc}~=~m_a\,(1~+~(v_{\rm lab}~+~\vesc)^2/2)$. 
To simplify this discussion we assume that the noise distribution in $\omega$ is flat, but this may not always be the case.

In Fig.~\ref{fig:axionstream} we demonstrate the impact of the S1 stream for a range of density fractions and dispersions 
(quoting 1-dimensional dispersion values for simplicity). 
We show the ratios of the minimum coupling measurable when the SHM+S1 model is assumed, 
relative to the SHM model alone. This result is independent of both the axion mass and whether the experiment is broadband or resonant. 
Since a $\chi^2$ or log-likelihood ratio test statistic depends on the square of the speed distribution, 
for very cold streams the sensitivity to DM significantly improves when a stream is present, even when its density fraction is low. 

For larger values of the stream dispersion, including the wide dispersion S1 stream, Fig.~\ref{fig:axionstream} shows that the DM sensitivity is actually slightly suppressed compared to a halo with no stream. The dispersion here again is inferred from the stellar dispersion so therefore may be overestimated cf.~Sec.~\ref{sec:halomodel}. The suppression in sensitivity for S1 is nonetheless minor and statistical fluctuations in real data will likely have a bigger impact on the sensitivity. The enhancement for a wide dispersion stream when its density comprises the majority of the local DM occurs because the stream distribution by itself is sharper than the SHM.

In contrast to experiments searching for WIMPs, the fact that the DM associated with S1 moves with a higher lab speed
makes it slightly harder to detect than streams that have a smaller speed (such as the Sagittarius stream).
This is because when binned in frequency, the relation $\Delta \omega \propto v \Delta v$ holds. 
Since $\Delta \omega$ is constant as it is fixed by the experimental design, a feature in the data
of a given width in speed $\Delta v$ will be spread over more frequency bins when centred at higher~$v$. 
This means that the signal-to-noise ratio in each frequency bin will be slightly lower, and the high speed
stream will be slightly harder to detect.

 In Fig.~\ref{fig:axionlimits} we demonstrate the same change in sensitivity as in Fig.~\ref{fig:axionstream} 
 but now in the more familiar $g_{a\gamma}-m_a$ plane
for the ADMX, MADMAX and ABRACADABRA haloscopes described previously.\footnote{Only ADMX has published limits~\cite{Du:2018uak} in this plane so here, we show projections for the next few upgrades.} For clarity, we only show the change in sensitivity for a stream dispersion of $5$~km~s$^{-1}$, where the sensitivity increase is relatively large and so can easily be observed on the log-scale plot.

\subsection{Measuring properties of S1}
Once a detection of axion DM has been made, a detailed measurement of the frequency dependence of the signal can be performed.
We display examples of the power spectra in Fig.~\ref{fig:axion_spectrum} for two values of the stream dispersion.
Comparing the left hand panel with the right hand panel, which has a colder dispersion velocity, we see clearly the importance of $\sigma_{\rm str}$ for detecting the stream. Power spectra spectra like the ones shown in Fig.~\ref{fig:axion_spectrum}
could be straightforwardly obtained in a resonant device by simply running the experiment at a single resonant frequency, with no need to tune. 
The frequency resolution is given by the inverse of the integration time for a single time-stream sample but once the axion is detected this duration can be made much longer. The potential signal-to-noise that can be obtained quickly is extremely high. With resonant methods, covering the mass ranges shown in Fig.~\ref{fig:axionlimits} requires up to $10^6$ mass points to be scanned over a time of several years. Post-detection we need only one. In fact the detection of the axion would only require the peak be located above the noise to some decent statistical level, like 3$\sigma$, at one mass point. This means that the second phase fixed-frequency experiment with a duration of the same timescale would effectively have a 3$\sigma$ signal $10^6$ times stronger. This would allow incredibly fine-grained studies of the local axion field, and the for the S1 stream to be measured down to even lower density fractions than shown here~\cite{Foster:2017hbq,OHare:2017yze}.

Since the axion signal is directly given by the speed distribution, the S1 stream can simply be fit from the power spectrum. This kind of analysis has already been explored in detail in Refs.~\cite{OHare:2017yze,Foster:2017hbq,Knirck:2018knd}.
 The annual modulation phase and amplitude enables very accurate measurements of the stream velocity. Even though S1 is much wider and less prominent in the power spectrum than the examples used in previous studies (which are much closer to the right hand side example of Fig.~\ref{fig:axion_spectrum}), a high signal-to-noise ratio can be obtained once the axion mass (and therefore frequency) is known and properties of S1 should be easily extracted from the data. In stark contrast to the WIMP detectors discussed in Secs.~\ref{sec:xenon} and~\ref{sec:directional}, the prospects for axion astronomy are rather extraordinary.

\section{Summary}\label{sec:summary}

The S1 stream, recently discovered in the SDSS-{\it Gaia} dataset, 
hits the Solar system almost head-on in a low inclination, counter rotating orbit 
(cf.\ Figs.~\ref{fig:S1pic} and~\ref{fig:S1picnew}). 
The DM particles associated with S1 have a much larger velocity in the 
laboratory frame compared with the DM particles from the rest of the halo (cf.\ Fig.~\ref{fig:fv}). The S1 stream is more akin to a DM hurricane than merely a wind.

By matching the spatial and kinematic properties of the S1 stream, 
its progenitor is believed to be a dwarf galaxy of mass $\approx 10^{10} M_\odot$ in stars and DM.
The present-day DM content of S1 is not known, but is expected to be significant. 
In this work, we have studied the effects of the S1 stream on WIMP and axion direct detection experiments 
while remaining agnostic as to the precise fraction of the local DM density contributed to by the S1 stream.

We have first examined the prospects for the detection of S1 in upcoming multi-ton xenon-based WIMP detectors. 
In this case, we find that S1 can only be unambiguously detected for WIMPs with masses in the approximate 
range between~$5$ and~$25$~GeV, when the stream density comprises an $\mathcal{O}(10\%)$
fraction of the local density (cf.\ Fig.~\ref{fig:streamDL_xenon}).
Xenon detectors are ultimately limited because much of the kinematic information is lost through nuclear scattering. Over much of the WIMP mass parameter space, the S1 and SHM recoil spectra look extremely similar 
(cf.~Figs.~\ref{fig:xenon-rate} and~\ref{fig:annual-modulation}).

Next, we examined the prospects for CYGNUS, a future \emph{directional} detector. 
In this case, the detection prospects are more promising since S1 creates characteristic 
ring-like features in the angular recoil spectrum (cf.~Fig.~\ref{fig:directional}). 
We find that S1 enhances the promise of directional detectors because it increases the degree of anisotropy in the WIMP signal. For spin-independent WIMP-nucleus scattering, CYGNUS is complementary to future xenon detectors since it extends the sensitivity to S1 down to approximately 0.8 GeV. For spin-dependent scattering, CYGNUS has the potential to detect S1 over the full WIMP mass range (cf.~Fig.~\ref{fig:streamDL_cygnus}).

Finally, we considered detecting S1 if DM was not a WIMP but instead an axion. 
Relative to nuclear recoil experiments, axion haloscopes have unmatched sensitivity to the local DM velocity distribution.  The S1 stream could have two observable effects. Firstly, depending on the dispersion, it could improve the prospects for detecting axion DM (cf.~Figs.~\ref{fig:axionstream} and~\ref{fig:axionlimits}). Secondly, after the axion mass has been identified, properties of the stream can essentially be read from the 
power spectrum of an axion-induced electromagnetic signal time-stream (cf.~Fig.~\ref{fig:axion_spectrum}). The detection of the S1 stream --- or any other DM substructure that may be present in our local halo --- has truly excellent detection prospects if the DM in our galaxy is made up of axions.

\acknowledgments
CAJO thanks Sven Vahsen for useful discussions about CYGNUS and VB thanks Matthew Buckley for interesting discussions.
 CAJO is supported by the grant FPA2015-65745-P from the Spanish MINECO and European FEDER. 
CM is supported by the Science and Technology Facilities Council (STFC) Grant ST/N004663/1. 
GCM thanks the Boustany Foundation, Cambridge Commonwealth, European \& International Trust and Isaac Newton Studentship for their support of his work. 
The research leading to these results has received funding from the European Research Council under the European Union's Seventh Framework Programme (FP/2007-2013) / ERC Grant Agreement n.\ 308024.

\maketitle
\flushbottom

\bibliographystyle{apsrev4-1}
\bibliography{S1stream}

\end{document}